\documentclass[%
	a4paper,%
	oneside,%
	11pt,%
]{scrartcl}

\usepackage[latin1]{inputenc}
\usepackage{amsmath}
\usepackage{amssymb}
\usepackage[T1]{fontenc} 
\usepackage{graphicx}
\usepackage{caption}
\usepackage{subcaption}
\usepackage{textcomp}
\usepackage{gensymb}
\usepackage{url}
\usepackage{xcolor}
\definecolor{darkblue}{rgb}{0,0,.5}
\definecolor{darkgreen}{rgb}{0,0.5,0}
\definecolor{darkred}{rgb}{0.5,0,0}
\usepackage[%
        hyperfootnotes=false,
    plainpages=false,
    pdfpagelabels,
    colorlinks=true,
    breaklinks=true,
    linkcolor=darkblue,
    menucolor=darkblue,
    urlcolor=darkred,
    citecolor=darkgreen,
  citebordercolor=darkgreen
]{hyperref}
\usepackage{natbib}
\usepackage{textcomp}
\usepackage{booktabs}

\graphicspath{{figures/}}
\newcommand{\fracpd}[2]{\frac{\partial #1}{\partial #2}} 
\newcommand{\fracd}[2]{\frac{\text{d} #1}{\text{d} #2}}

\usepackage{soul}

\usepackage{cancel}

\begin{document} 
\title{Ray-tracing in pseudo-complex General Relativity}
\author{T. Sch\"onenbach$^1$, G. Caspar$^1$, P. O. Hess$^{1,2,3}$, \\
  T. Boller$^4$, A. M\"uller$^5$, M. Sch\"afer$^1$ and W. Greiner$^1$ \\
  {\small\it $^1$Frankfurt Institute for Advanced Studies, Johann Wolfgang Goethe Universit\"at,} \\
  {\small\it Ruth-Moufang-Str. 1, D-60438 Frankfurt am Main, Germany} \\
  {\small\it $^2$GSI Helmholtzzentrum f\"uer Schwerionenforschung GmbH,}\\
  {\small\it Max-Planck-Str. 1, D-64291 Darmstadt, Germany}\\
  {\small\it $^3$ Instituto de Ciencias Nucleares, UNAM, Circuito Exterior, C.U.,} \\
  {\small\it A.P. 70-543, 04510 M\'exico, D.F., Mexico} \\
  {\small\it $^4$ Max-Planck Institute for Extraterrestial Physics, Giessenbachstrasse,} \\
  {\small\it D-85748 Garching, Germany}\\
  {\small\it $^5$ Excellence Cluster Universe, TU M\"unchen, Boltzmannstrasse 2,} \\
  {\small\it D-85748 Garching, Germany}
}

\maketitle

\begin{center}
  Accepted 2014 April 28.  Received 2014 April 28; in original form 2013 December 4
\end{center}

\begin{abstract}
Motivated by possible observations of the black hole candidate in the center of our
          galaxy \citep{gillessen, Eisenhauer_GRAVITY} and the galaxy M87 \citep{doeleman, falcke}, ray-tracing methods are 
          applied to both standard General Relativity (GR) and 
          a recently proposed extension, the pseudo-complex General Relativity (pc-GR). The 
          correction terms due to the investigated pc-GR model lead to slower orbital motions
          close to massive objects. Also the concept of an innermost stable circular orbit (ISCO)
          is modified for the pc-GR model, allowing particles to get closer to the central object
          for most values of the spin parameter $a$ than in GR.
          Thus, the accretion disk, surrounding a massive object,
          is brighter in pc-GR than in GR. Iron K$\alpha$ emission line profiles are also
          calculated as those are good observables for regions of strong gravity. Differences 
          between the two theories are pointed out. 
\end{abstract}

\section{Introduction}
\label{sec:introduction}
  Taking a picture of a black hole is not possible as long as an ambient light source is missing. 
  However, we can image a black hole and its strong gravitational effects by following light 
  rays coming from a source near the black hole. A powerful standard technique is called 
  \emph{ray-tracing}. The basic idea is to follow light rays (on null geodesics) in a 
  curved background spacetime from their point of emission, e.g. in an accretion disk, around a massive 
  object\footnote{Technically this is not correct. It is computationally less expensive to follow 
  light rays from an observers screen back to their point of emission.}. In this way one can create
  an image of the black holes direct neighbourhood. There are numerous groups using 
  ray-tracing for this purpose, see, e.g. \cite{fanton,muller1,gyoto,bambi}. Aside from an image 
  of the black hole it is also possible to calculate emission line profiles using the same 
  technique, but adding in a second step the evaluation of an integral for the spectral flux. This
  is of particular interest as the emission profile of, e.g. the iron K$\alpha$ line
  is one of the few good observables in regions with strong gravitation. 

  In the near future it will be possible to resolve the central massive object Sagittarius A* in
  the centre of our galaxy and the one in M87 with the planned Event Horizon Telescope \citep{doeleman,falcke}.
  This offers a great opportunity to test General Relativity and its predictions. 

  Predicting the expected picture from theory gets even more important, noting that during 2013/2014
  a gas cloud approaches close to the centre of our  galaxy \citep{gillessen} 
  and probably a portion of it may become part of an accretion disk. Once formed, we assume it also may 
  exhibit hot spots, seen as \emph{Quasi Periodic Oscillations} (QPOs) \citep{QPO} with the possibility to 
  measure the iron K$\alpha$ line. This gives the chance to test a theory, measuring the
  periodicity of the QPO and simultaneously the redshift.

  Recently, in \cite{pc-gr-2009,pc-gr-2012}, a pseudo-complex extension to General Relativity (pc-GR) was
  proposed, which adds to the usual coupling of mass to the geometry of space as a new ingredient
  the presence of a dark energy fluid with negative energy density.  The resulting changes
  to Einstein's equations could also be obtained by introducing a non vanishing energy momentum tensor in
  standard GR but arise more naturally when using a pseudo-complex description.
  By another group, in \cite{visser} the coupling of the mass to the local quantum property of vacuum fluctuations
  was investigated, applying semi-classical Quantum Mechanics, where the decline of the energy
  density is dominated by a $1/r^6$ behaviour. However, in \cite{visser} no recoupling to the metric
  was considered. In pc-GR the recoupling to the metric is automatically included and the 
  energy density is modelled to decline as $1/r^5$, however there is no microscopical description
  for the dark energy yet. The fall-off of order $1/r^5$  can neither be noted yet by solar system 
  experiments \citep{solar}, nor in systems of two orbiting neutron stars \citep{hulse-taylor}.
  A model description of the Hulse-Taylor-Binary, including pc-GR terms,
  showed that corrections become significant twelve orders of magnitude beyond current accuracy.
  Other models concerning the physics of neutron stars are currently prepared for publication \citep{neutronstars}.
  \par
  The effects of the dark energy become important near the Schwarzschild radius of a compact object
  towards smaller radial distances. 
  A parameter $B=bm^3$ is introduced, which defines the coupling of the mass with the 
  vacuum fluctuations. In contrast to the work by \cite{visser} the coupling of the mass 
  to vacuum fluctuations on a macroscopical level, described with the parameter $B$, allows to
  include alterations to the metric.
  A downside is that this modification yet lacks a complete microscopical description, thus one 
  can see the work by \cite{visser} and pc-GR as complementary.
  In \cite{pc-gr-2012} investigations showed that a value of $B> (64/27) m^3$ leads to a metric 
  with {\it no event horizons}. Thus, an external observer can in principle still look inside, 
  though a large redshift will make the grey star look like a black hole. In the 
  following we will use the critical value $B= (64/27) m^3$ if not otherwise stated.

  In \cite{astro-1} several predictions were made, related to the motion of a test particle in a 
  circular orbit around a massive compact object (labelled there as a \emph{grey star}), which is
  relevant for the observation of a QPO and the redshift. One distinct feature is that at a certain
  distance in pc-GR the orbital frequency shows a maximum, from which it decreases again toward 
  lower radii, allowing near the surface of the star a low orbital frequency correlated with a 
  large redshift. This will affect the spectrum as seen by an observer at a large distance. Thus, 
  it is of interest to know how the accretion disk would look like by using pc-GR. In addition to
  the usual assumptions made for modelling accretion disks, e.g. in \cite{pagethorne}, we assume
  that the coupling of the dark energy to the matter of the disk to be negligible compared to 
  the coupling to the central object. This is justified in the same way as one usually neglects the
  mass of the disks material in comparison to the central object.

  In the following we will first briefly review the theoretical background on the methods used,
  where we will also discuss the two models we used to describe accretion disks.
  After that we will present results obtained with the open source ray-tracing code 
  \textsc{Gyoto}\footnote{\textsc{Gyoto} is obtainable at \url{http://gyoto.obspm.fr/}.} \citep{gyoto} for 
  the simulation of disk images and emission line profiles.


\section{Theoretical Background}
\label{sec:theory}
  We will use the Boyer-Lindquist coordinates of the Kerr metric and its pseudo-complex equivalent, 
  which we will write in a slightly different form\footnote{Here we use the convention 
  $a =\frac{\kappa J}{m}$ instead of $a= -\frac{\kappa J}{m}$, where $J$ is the angular 
  momentum of the central massive object, and signature (-,+,+,+) in contrast to previous work in 
  \cite{pc-gr-2012, astro-1}.} than in \cite{pc-gr-2012} as
  \begin{align}
    \label{eq:kerr_1}
    g_{00} &=  -\left(1-\frac{\psi}{\Sigma} \right)\quad, \nonumber\\
    g_{11} &=  \frac{\Sigma}{\Delta}\quad, \nonumber\\
    g_{22} &=  \Sigma\quad, \nonumber\\
    g_{33} &= \left(
    (r^{2} + a^{2})+\frac{a^{2}\psi}{\Sigma}\sin^2\theta
    \right)\sin^2\theta \quad,\nonumber\\
    g_{03} &=-a\frac{\psi}{\Sigma}\sin^2\theta\quad, 
  \end{align}
  with
  \begin{align}
    \Sigma &= r^{2}+ a^{2}\cos^{2}\theta\quad ,\nonumber\\
    \Delta &= r^{2}+a^{2} - \psi\quad ,\nonumber\\
    \psi &= 2mr-\frac{B}{2r}\quad .
    \label{eq:pc_modified_BL}
  \end{align}
  Here $m = \kappa M$ is the gravitational radius of a 
  massive object, $M$ is its mass, $a$ is a measure for the specific angular momentum or spin
  of this object and $\kappa$ is the gravitational constant. In addition we set the speed of light $c$ to one.
  One can easily see that this metric differs from the standard Kerr metric only in the use of the
  function $\psi$ which reduces to $2mr$ in the limit $B=0$. Bearing this in mind one can simply 
  follow the derivation of the Lagrange equations given, e.g. in \cite{levin:2008} (which are the 
  basis for the implementation in \textsc{Gyoto}) and modify the occurrences of the Boyer-Lindquist 
  $\Delta$-function and the new introduced $\psi$. \\
  To derive the desired equations we exploit all conserved quantities along geodesics which are
  the test particle's mass $m_0$, the energy at infinity $E$, the angular momentum $L_z$ and
  the Carter constant $Q$ \citep{levin:2008, carter}. The usual way to proceed from this, is to
  follow \cite{carter} and demand separability of Hamilton's principal function
   \begin{equation}
     S = - \frac{1}{2}  \lambda - E t + L_z \phi +S_\theta +S_r \quad.
     \label{eq:hamilton_principal}
     \end{equation}
   Here $\lambda$ is an affine parameter and $S_r$ and $S_\theta$ are functions of $r$ and $\theta$,
   respectively. Demanding separability now for equation (\ref{eq:hamilton_principal}) leads 
   \cite{carter} to
     \begin{equation}
       \left(\fracd{S_r}{r}\right)^2 = \frac{R}{\Delta^2} \quad \text{and} \quad 
       \left(\fracd{S_\theta}{\theta}\right)^2 = \Theta \quad ,
       \label{eq:S_r_theta}
     \end{equation}
    with the auxilliary functions
  \begin{align}
    R(r) &:= \left[(r^2+a^2)E  - aL_z \right]^2   \notag \\
         &\phantom{:=} - \Delta\left[Q+ (aE-L_z)^2 +m_0^2 r^2 \right]\quad, \notag \\
    \Theta(\theta) &:= Q - \left[\frac{L_z^2}{\sin^2\theta} -a^2E^2 + m_0^2 a^2 \right]\cos^2\theta\quad.
    \label{eq:bigRT}
  \end{align}
  Taking these together with
  \begin{equation}
    \dot{x}^\mu = g^{\mu\alpha} p_\alpha = g^{\mu\alpha} \fracpd{S}{x^\alpha}      
  \end{equation} 
  leads to a  set of first order equations of motion in the coordinates 
  \begin{align}
    \dot{t} &= \frac{1}{\Sigma \Delta} \left\{ \left[ \left( r^2 +a^2 \right)^2
                + a^2 \Delta \sin^2\theta \right] E - a \psi L_z \right\} \notag \\
    \dot{r} &= \pm \frac{\sqrt{R}}{\Sigma} \notag \\
    \dot{\theta} &= \pm \frac{\sqrt{\Theta}}{\Sigma} \notag \\
    \dot{\phi} &= \frac{1}{\Sigma \Delta} \left[ \left( \frac{\Delta}{\sin^2\theta}
                 - a^2\right) L_z + a \psi E\right] \quad,
    \label{eq:eom_ray-tracing_std}
   \end{align}
   where the dot represents the derivative with respect to the proper time $\tau$.

  \cite{levin:2008} however argue, that those equations contain an ambiguity in the sign for the 
  radial and azimuthal velocities. In addition to that, using Hamilton's principle to
  get the geodesics leads to the integral equation \citep{carter}
  \begin{equation}
   \int\limits_{r_\text{em}}^{r_\text{obs}} \frac{\text{d}r}{\sqrt{R}} = 
       \int\limits_{\theta_\text{em}}^{\theta_\text{obs}} \frac{\text{d}\theta}{\sqrt{\Theta}} \quad.
  \end{equation}  
  To solve this equation \cite{fanton, muller1} make use of the fact that $R$ is a fourth order 
  polynomial in $r$. This is not possible anymore with the pseudo-complex correction terms as 
  the order of $R$ increases. \\
  Luckily the equations used in \textsc{Gyoto} are based on the use of different equations of
  motion derived by \cite{levin:2008}.  Therefore we follow 
  \cite{levin:2008} who make use of the Hamiltonian formulation in addition to
  the separability of Hamilton's principal function. The canonical
  4-momentum of a particle is given as
  \begin{equation}
    p^\mu := \dot{x}^\mu \quad,
  \end{equation}
  when we write the Lagrangian as
  \begin{equation}
  \mathcal{L} = \frac{1}{2} g_{\mu\nu} \dot{x}^\mu \dot{x}^\nu \quad.
  \end{equation}
  Given explicitly
  in their covariant form the momenta are \citep{levin:2008}
  \begin{align}
    p_0 &= - \left( 1- \frac{\psi}{\Sigma} \right) \dot{t} - \frac{\psi a \sin^2\theta}{\Sigma} 
             \dot{\phi}\quad,  \notag \\
    p_1 &= \frac{\Sigma}{\Delta} \dot{r}\quad,  \notag \\
    p_2 &= \Sigma \dot{\theta}\quad,  \notag \\
    p_3 &= \sin^2\theta \left( r^2 + a^2 + \frac{\psi a^2  \sin^2\theta}{\Sigma} \right)
             \dot{\phi} - \frac{\psi a \sin^2\theta}{\Sigma} \dot{t} \quad. 
    \label{eq:4momenta}    
  \end{align}
  After a short calculation the Hamiltonian $\mathcal{H} = p_\mu \dot{x}^\mu - \mathcal{L} 
  = \frac{1}{2} g^{\mu\nu} p_\mu p_\nu$ can be rewritten as \citep{levin:2008}
  \begin{equation}
    \mathcal{H} = \frac{\Delta}{2\Sigma} p_1^2 + \frac{1}{2 \Sigma} p_2^2 
                  - \frac{R(r) + \Delta\Theta(\theta)}{2 \Delta \Sigma} - \frac{m_0}{2} \quad .
  \end{equation}
  The momenta associated with time $t$ and azimuth $\phi$ are conserved and can be identified with
  the energy at infinity $p_0 = -E$ and the angular momentum $p_3 = L_z$,
  respectively~\citep{carter}. \\
  Now Hamilton's equations $\dot{x}_\mu = \fracpd{\mathcal{H}}{p_\mu}$ and $\dot{p}_\mu = 
  - \fracpd{\mathcal{H}}{x_\mu}$ yield the wanted equations of motion\footnote{Each occurrence of
  $p_0$ and $p_3$ here is already replaced by the constants of motion $-E$ and $L_z$, respectively.} 
  \citep{gyoto,levin:2008}
  \begin{align}
    \dot{t} &= \frac{1}{2 \Delta\Sigma} \fracpd{}{E} (R + \Delta\Theta)\quad,  \notag \\
    \dot{r} &= \frac{\Delta}{\Sigma} p_1 \quad, \notag \\
    \dot{\theta} &= \frac{1}{\Sigma} p_2 \quad, \notag \\
    \dot{\phi} &= - \frac{1}{2\Delta\Sigma} \fracpd{}{L_z} (R + \Delta\Theta) \quad, \notag \\
    \dot{p}_0 &= 0 \quad, \notag \\
    \dot{p}_1 &= - \left(\frac{\Delta}{2\Sigma}\right)_{|r} p_1^2 - \left(\frac{1}{2\Sigma}\right)_{|r} p_2^2
                 + \left( \frac{R +\Delta \Theta}{2\Delta \Sigma} \right)_{|r} \quad, \notag \\
    \dot{p}_2 &= - \left(\frac{\Delta}{2\Sigma}\right)_{|\theta} p_1^2 
                      - \left(\frac{1}{2\Sigma}\right)_{|\theta} p_2^2
                      + \left( \frac{R +\Delta \Theta}{2\Delta \Sigma} \right)_{|\theta} \quad, \notag \\
    \dot{p}_3 &= 0 \quad.
    \label{eq:evolutionequations}
  \end{align}
  Here ${}_{|r}$ and ${}_{|\theta}$ stand for the partial derivatives with respect to $r$ and $\theta$.

  In addition to the modification of the metric and thus the evolution equations one has to modify
  the orbital frequency of particles around a compact object. This has been done in \cite{astro-1}
  with the resulting frequencies 
  \begin{equation}
    \omega_{\pm}=\frac{1}{a\mp\sqrt{\frac{2r}{h(r)}}}\quad,
    \label{eq:opm}
  \end{equation}
  where $\omega_-$ describes prograde motion and $h(r) = \frac{2m}{r^2} - \frac{3B}{2r^4}$.
  Equation (\ref{eq:opm}) reduces to the well known $\omega_\pm = \frac{1}{a \mp 
  \sqrt{\frac{r^3}{m}}}$ for $B=0$. 
  \\Finally the concept of an innermost stable circular orbit (ISCO) has to be revised, as the 
  pc-equivalent of the Kerr metric only shows an ISCO for some values of the spin parameter
  $a$. For values of $a$ greater than $0.416m$ and $B=\frac{64}{27} m^3$ there is no region of 
  unstable orbits anymore \citep{astro-1}. 

  After including all those changes due to correction terms of the pseudo-complex equivalent of the
  Kerr metric one can straightforwardly adapt the calculations done in \textsc{Gyoto}. The adapted version
  will be published online soon.
  \subsection*{Obtaining observables}  
    \label{sec:observables}
    After setting the stage for geodesic evolution used for ray-tracing we will briefly discuss 
    physical observables used in ray-tracing studies. First of all let us note, that we will
    focus on ray-tracing of null-geodesics and thus our observables are of radiative nature. The
    first quantity of interest is the intensity of the radiation. The intensity of radiation
    emitted between a point $s_0$ and the position $s$ in the emitters frame is given 
    by~\citep{gyoto,rybicki}
    \begin{equation}
      I_\nu(s) = \int\limits_{s_0}^s\exp\left(-\int\limits_{s'}^s \alpha_\nu(s'')\text{d}s''\right)
                 j_\nu(s') \text{d}s' \quad. 
      \label{eq:intensity}
    \end{equation}
    Here $\alpha_\nu$ is the absorption coefficient and $j_\nu$ the emission coefficient in the
    comoving frame. \\
    Using the invariant intensity $\mathcal{I} = I_\nu/\nu^3$ \citep{mtw} one gets the observed 
    intensity via
    \begin{equation}
      I_{\nu_\text{obs}} = g^3 I_{\nu_\text{em}} \quad,
    \end{equation}
    where we introduced {the relativistic generalised redshift factor
    $g := \frac{\nu_\text{obs}}{\nu_\text{em}}$. The quantity observed however is the flux $F$
    which is given by 
    \begin{equation}
      \text{d}F_{\nu_\text{obs}} = I_{\nu_\text{obs}} \cos\vartheta \text{d} \Omega \quad,
    \end{equation}
    where $\vartheta$ describes the angle between the normal of the observers screen and the
    direction of incidence and $\Omega$ gives the solid angle in which the observer sees the 
    light source \citep{gyoto}. \\ \par
    In the following we will consider two special cases for the intensity. First the emission 
    line in an optically thick, geometrically thin accretion disk, which can be modelled 
    by~\citep{fanton,gyoto}
    \begin{equation}
      I_\nu \propto \delta(\nu_\text{em} - \nu_\text{line}) \varepsilon(r) \quad,
      \label{eq:int_powerlaw}
    \end{equation}
    where the radial emissivity $\varepsilon(r)$ is given by a power law 
    \begin{equation}
      \varepsilon(r) \propto r^{-\alpha} \quad,
      \label{eq:powerlaw_line}
    \end{equation}
    with $\alpha$ being the single power law index. 

    The second emission model we consider is a geometrically thin, infinite accretion disk first
    modelled by \cite{pagethorne}. The intensity profile here is strongly 
    dependent on the used metric and thus some modifications have to be done. Fortunately most of
    the results of \cite{pagethorne} can be inherited and only at the end one has to insert the 
    modified metric. Equation (12) in \cite{pagethorne} 
    \begin{equation}
      f = - \omega_{|r}(E-\omega L_z)^{-2} \int\limits_{r_{ms}}^r 
            (E-\omega L_z) {L_z}_{|r} \text{d}r
      \label{eq:f_pagethorne}
    \end{equation}
    builds the core for the computation of the flux \citep{pagethorne} 
    \begin{equation}
      F = \frac{\dot{M}_0}{4\pi \sqrt{-g}} f \quad.
      \label{flux_pagethorne} 
    \end{equation}
    Assuming $\dot{M}_0 = 1$ as in \cite{gyoto} and observing that the determinant of the metric 
    $\sqrt{-g}$ is the same for both GR and pc-GR, we see that the only difference in the flux lies
    in the function $f$ given by equation~(\ref{eq:f_pagethorne}).  \\
    In addition to the assumptions made by \cite{pagethorne} we have to include the 
    assumption that the stresses inside the disk carry angular momentum and energy from faster to
    slower rotating parts of the disk. In the case of standard GR this assumption means that
    energy and angular momentum get transported outwards. In pc-GR equation (\ref{eq:f_pagethorne})
    then has to be modified to
    \begin{equation}
      f = - \omega_{|r}(E-\omega L_z)^{-2} \int\limits_{r_{\omega_\text{max}}}^r 
            (E-\omega L_z) {L_z}_{|r} \text{d}r \quad ,
      \label{eq:f_pagethorne_mod}
    \end{equation} 
    where $\omega_\text{max}$ describes the orbit where the angular frequency $\omega$ has its
    maximum (this is the last stable orbit in standard GR). \\
    Equation~(\ref{eq:f_pagethorne_mod}) gives a concise way to write down the flux in the two regions
    ($r_\text{in}$ describes the inner edge of an accretion disk):
    \begin{enumerate}
     \item $r_{\omega_\text{max}} < r_\text{in} \leq r $: This is also the standard GR case, where $\omega_{|r} < 0$ and 
           the flux in equations~(\ref{eq:f_pagethorne}) and (\ref{eq:f_pagethorne_mod}) is positive.
     \item $r_\text{in} \leq r  < r_{\omega_\text{max}}$: Here $\omega_{|r} > 0$, but the upper integration limit in
           equation (\ref{eq:f_pagethorne_mod}) is smaller than the lower one. Thus there are overall two sign changes and
           the flux $f$ is positive again.
    \end{enumerate}
    Thus if we consider a disk whose inner radius is below $r_{\omega_\text{max}}$, which is the case in the pc-GR model 
    for $a>0.416m$, equation (\ref{eq:f_pagethorne_mod}) guarantees a positive flux function $f$.  \\ \par

All quantities $E, L_z, \omega$ 
    in (\ref{eq:f_pagethorne_mod}) were already computed in \cite{astro-1}. The angular frequency $\omega$ is given in 
    equation~(\ref{eq:opm}), $E$ and $L_z$ are given as\footnote{The change of signature
    and the sign of the spin parameter $a$ have to be kept in mind.} 
    \begin{align}
      L_z^2 &= 
      \frac{\left (g_{03} + \omega g_{33} \right )^2}{-g_{33}\omega^2 - 2g_{03}\omega - g_{00}} \quad, \notag\\
      E^2 &= 
      \frac{\left (g_{00} + \omega g_{03} \right )^2}{-g_{33}\omega^2 - 2g_{03}\omega - g_{00}} \quad .
      \label{eq:LandE2}
    \end{align}
    Unfortunately the derivatives of $E$ and $L_z$ become lengthy in pc-GR and
    the integral in equation~(\ref{eq:f_pagethorne}) has no analytic solution anymore. 
    Nevertheless it can be solved numerically and thus we are able to modify the original disk 
    model by \cite{pagethorne} to include pc-GR correction terms.


\section{Results}
\label{sec:results}
  As shown in \cite{astro-1} the concept of an ISCO is modified in the pc-GR model. For
  the following results we used as the inner radius for the disks in the pc-GR case the values depicted
  in Tab.~\ref{tab:rin_pc}. Values of $r_\text{in}$ for $a\leq 0.4m$ correspond to the modified last
  stable orbit.
  \begin{table}
   \begin{tabular}{cc}
      \toprule 
      Spin parameter $a$[m] & $r_\text{in}$[m] \\
       0.0 & 5.24392 \\
       0.1 & 4.82365 \\
       0.2 & 4.35976 \\
       0.3 & 3.81529 \\
       0.4 & 2.99911 \\
       0.5 and above & 1.334 \\
      \bottomrule
    \end{tabular}
    \caption{Values for the inner edge of the disks $r_\text{in}$ in pc-GR for the parameter 
    $B=64/27m^3$. }
    \label{tab:rin_pc}
  \end{table}
  The value of $r_\text{in}$ for values of $a$ above $0.416m$ is chosen slightly above the value 
  $r= (4/3)m$. For smaller radii equation~(\ref{eq:opm}) has no real solutions anymore  
  in the case of $B=(64/27)m^3$. The same also holds for general (not necessarily geodesic) circular orbits, where the
  time component $u^0 = \frac{1}{\sqrt{-g_{00} - 2 \omega g_{03} - \omega^2 g_{33}}}$ of the particles
  four-velocity also turns imaginary for radii below $r= (4/3)m$ in the case of $B=(64/27)m^3$.\\
  We assume that the compact
  massive object extends up to at least this radius. For all simulations however we did neglect any
  radiation from the compact object. This is a simplification which will be addressed in future
  works. \\
  The angular size of the compact object is also modified in the pc-GR case. It is proportional to
  the radius of the central object \citep{angular-size}, which varies in standard GR between 
  $1m$ and $2m$, leading to angular sizes of approximately $10-20~\mu as$ for Sagittarius A*. 
  The size of the central object in pc-GR is fixed at $r=(4/3)m$ in the limiting case 
  for $B=(64/27)m^3$ thus leading to an angular size of approximately $13~\mu as$. 
  \subsection{Images of an accretion disk}
    In Fig.~\ref{fig:accretiondisk_1} we show images of infinite geometrically thin accretion disks
    according to the model of \cite{pagethorne} (see section \ref{sec:observables})
    in certain scenarios. Shown is the bolometric intensity $I$[erg cm$^{-2}$s$^{-1}$ster$^{-1}$] 
    which is given by $I= \frac{1}{\pi} F$ \citep{gyoto}.
    To make differences comparable, we adjusted the scales for each value of the spin parameter $a$ 
    to match the scale for the pc-GR scenario. The plots of the Schwarzschild object ($a=0.0m$) and
    the first Kerr object ($a=0.3m$) use a linear scale whereas the plots for the other Kerr objects
    ($a=0.6m$ and $a=0.9m$ respectively) use a
    log scale for the intensity. This is a compromise between comparability between both theories 
    and visibility in each plot. One has to keep in mind, that scales remain constant for a 
    given spin parameter$a$ and change between different values for $a$. \par
    The overall behaviour is similar in GR and pc-GR. The most prominent difference is that the
    pc-GR images are brighter. 
    An explanation for this effect is the amount of energy which is released for particles
    moving to smaller radii. This energy is then transported via stresses to regions with lower angular
    velocity, thus making the disk overall brighter. In Fig.~\ref{fig:energy} we show this energy
    for particles on stable circular orbits. \\
    At first
    puzzling might be the fact that the fluxes differ significantly for radii above $10m$ although here
    the differences between the pc-GR and standard GR metric become negligible. However the flux $f$ in 
    equation~(\ref{eq:f_pagethorne_mod}) at any given radius $r$ depends on an integral over all radii
    starting from $r_{\omega_\text{max}}$ up to $r$. Thus the flux at relatively large radii
    is dependent on the behaviour of the energy at smaller radii, which differs significantly from standard GR.\\    
    It is important to stress that the difference in the flux between the standard GR and pc-GR 
    scenarios is thus also strongly dependent on the inner radius of the disk. This is due to the fact that the
    values for the energy too are strongly dependent on the radius, see figure~\ref{fig:energy_06}. In figure~\ref{fig:fluxfunction_b} we compare the pc-GR and GR case for the same inner radius. There is
    still a significant difference between both curves but not as strongly as in figure~\ref{fig:fluxfunction}.
    \par
    \begin{figure}
      \centering
      \begin{subfigure}{\columnwidth}
         \centering
         \includegraphics[width=.79\textwidth]{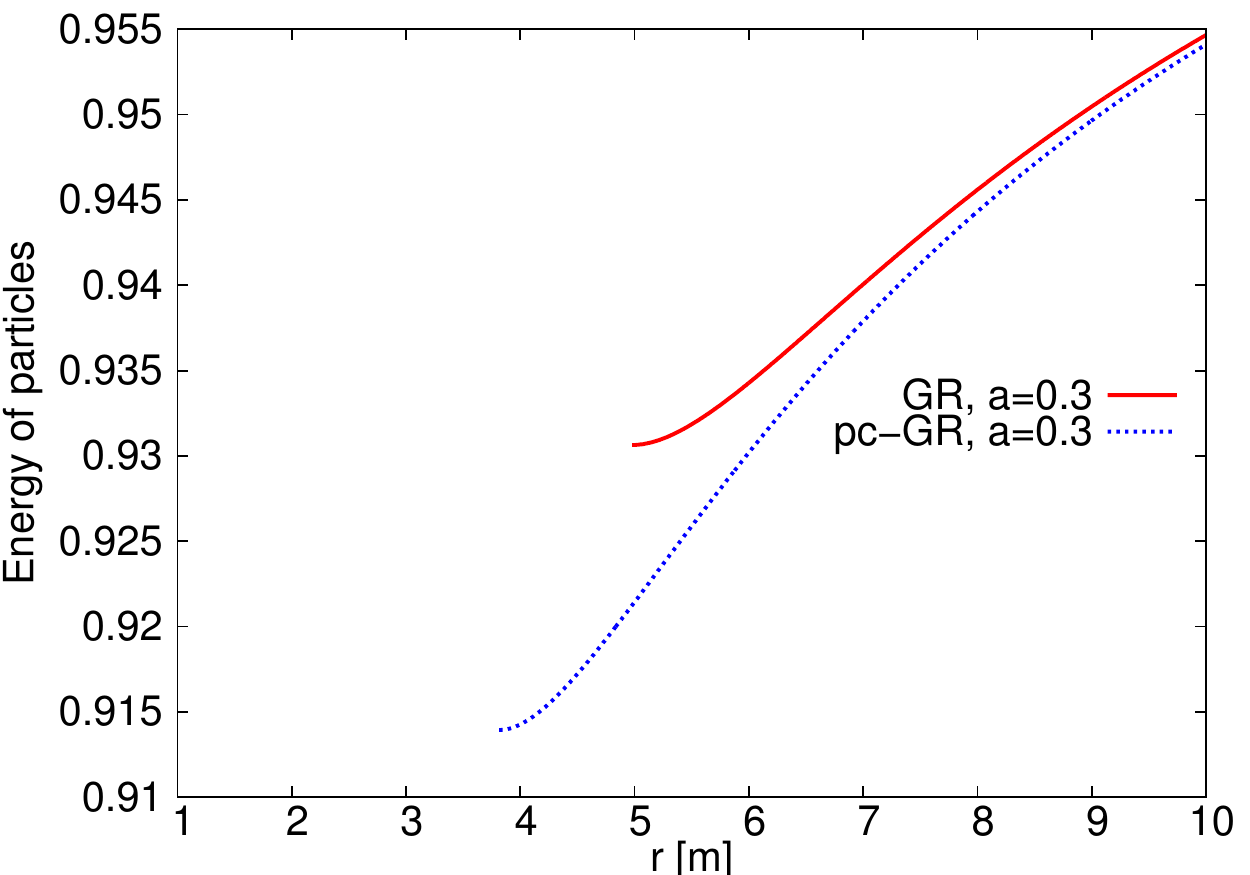}
         \caption{$a=0.3m$}
         \label{fig:energy_03}
      \end{subfigure}
      \begin{subfigure}{\columnwidth}
        \centering
        \includegraphics[width=.79\textwidth]{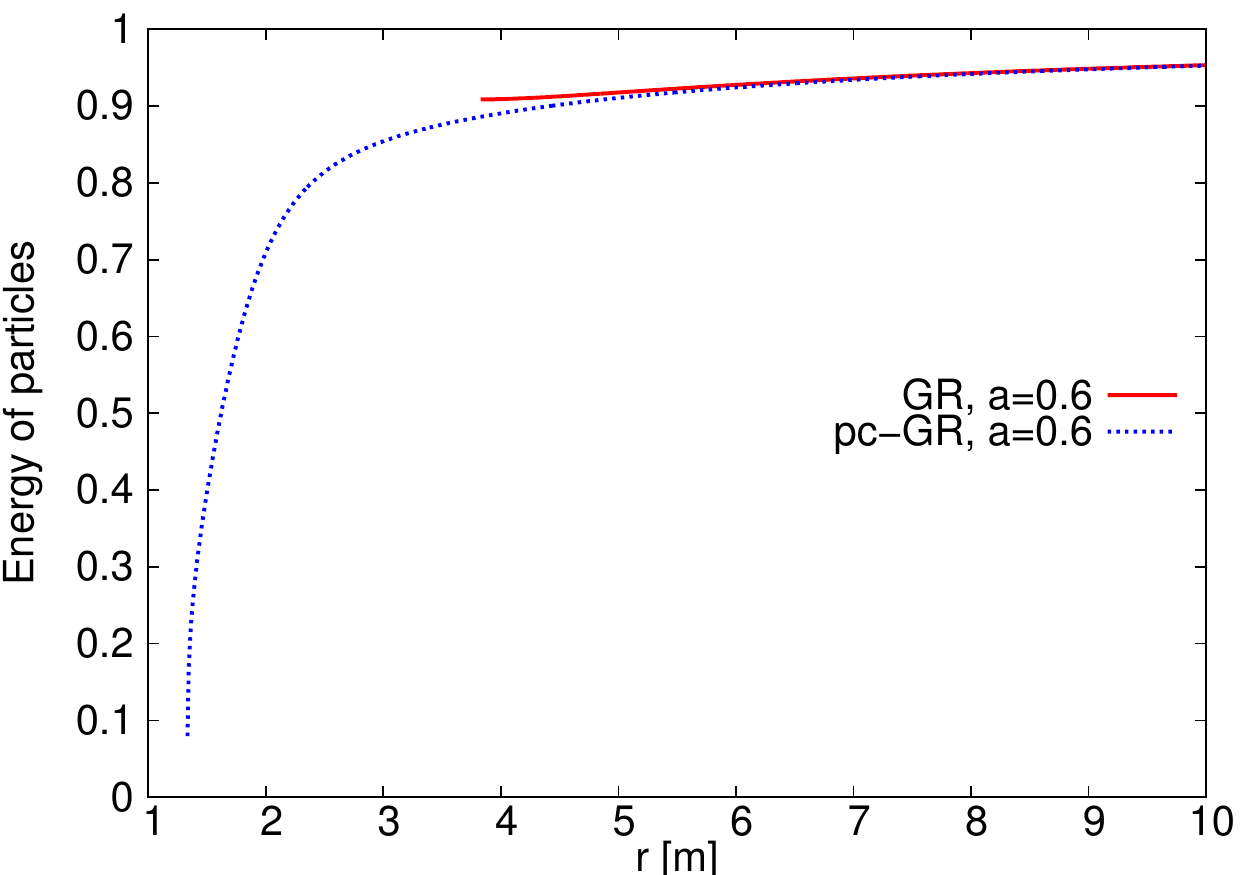}
        \caption{$a=0.6m$}
        \label{fig:energy_06}       
      \end{subfigure}
      \caption{Normalized energy of particles on stable prograde circular orbits. The pc-parameter $B$
               is set to the critical value of $(64/27)m^3$. In the pc-GR case more energy is released
               as particles move to smaller radii, where the ammount of released energy increases
               significantly in the case where no last stable orbit is present anymore. 
               The lines end at the last stable orbit or at $r=1.334m$, respectively.}
      \label{fig:energy}
    \end{figure}%
    The next significant difference to the standard disk model by \cite{pagethorne} is the occurrence
    of a dark ring in the case of $a \geq 0.416$. This ring appears in the pc-GR case due to the fact
    that the angular frequency of particles on stable orbits now has a maximum at 
    $r = r_{\omega_\text{max}} \approx 1.72 m$ \citep{astro-1} and
    the disks extend up to radii below $r_{\omega_\text{max}}$. At
    this point the flux function vanishes, see section \ref{sec:observables}. 
    Going further inside, the flux  increases again, which is a new feature of the pc-GR model. 
    This is the reason of the ring-like structure for $a>0.416m$. Note that the bright
    inner ring may be mistaken for second order effects although these do not appear as
    the disk extends up to the central object.
    \\
    In Fig.~\ref{fig:fluxfunction} we show the radial dependency of the flux function, 
    see equations~(\ref{eq:f_pagethorne}) and (\ref{eq:f_pagethorne_mod}).
    \begin{figure}
         \centering
         \begin{subfigure}{\columnwidth}
         \centering
         \includegraphics[width=.79\columnwidth]{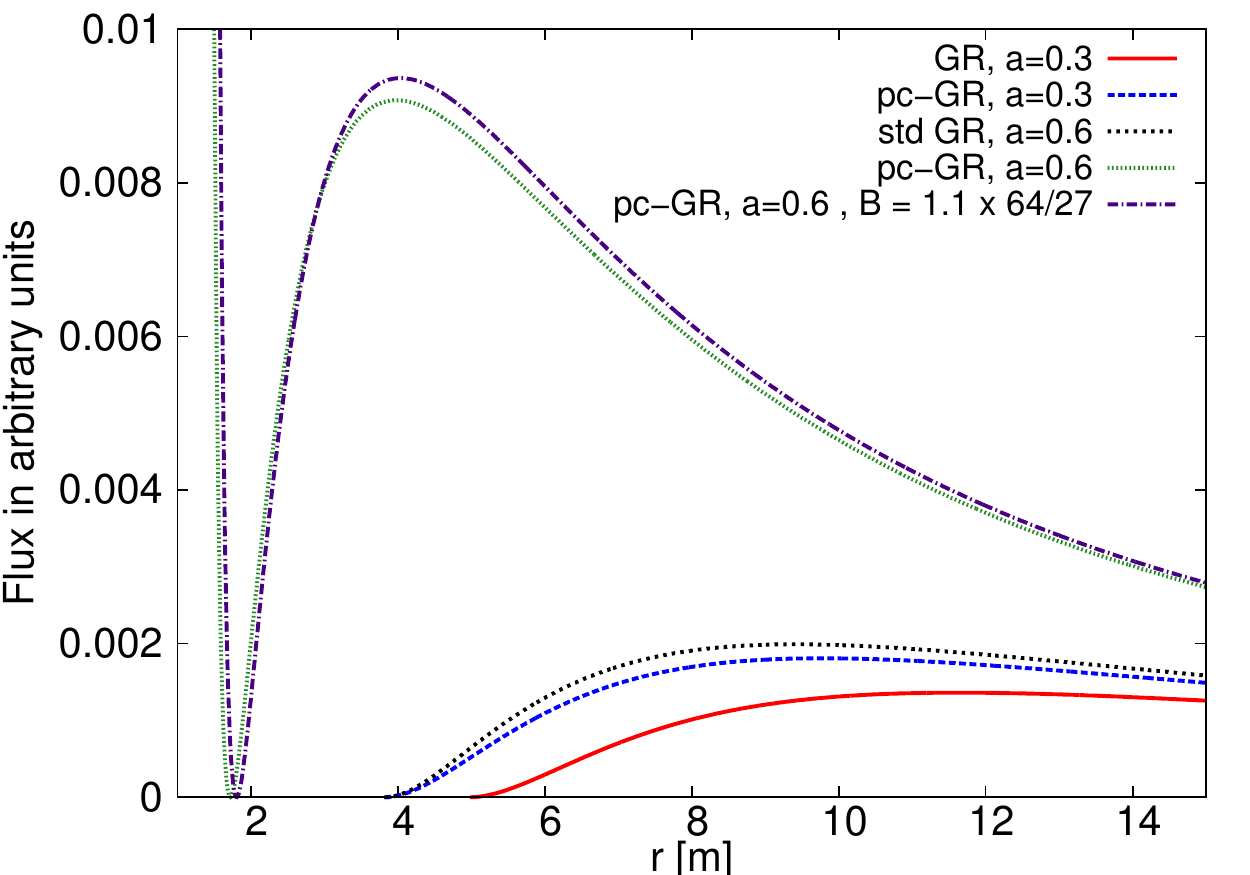}
         \caption{Flux function $f$ for varying spin parameter $a$ and 
                  inner edge of the akkretion disk. In the standard GR
                  case the ISCO is taken as inner radius, for the pc-GR case see table~\ref{tab:rin_pc}.}
         \label{fig:fluxfunction}         
         \end{subfigure}
         \begin{subfigure}{\columnwidth}
         \centering
         \includegraphics[width=.79\columnwidth]{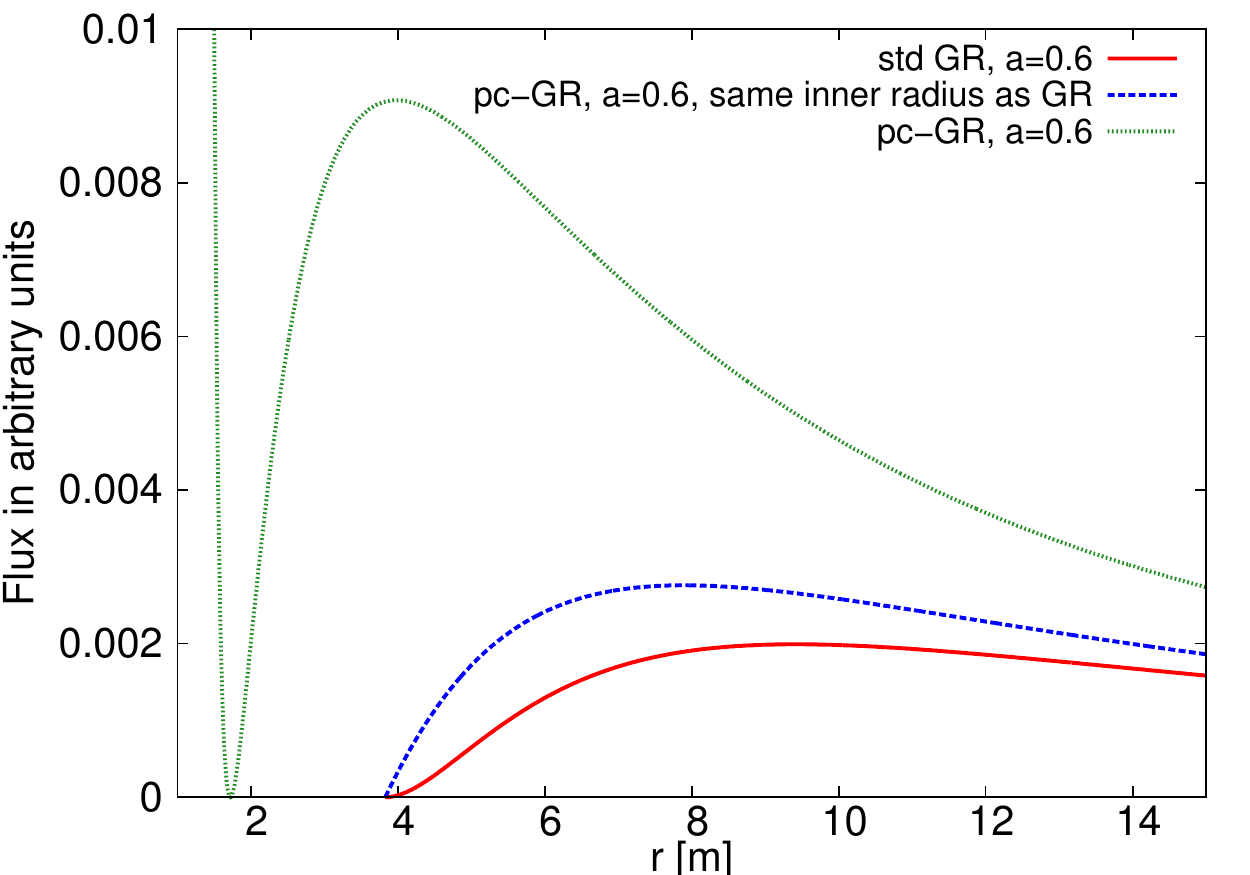}
         \caption{Dependence of the flux function $f$ on the inner radius of the disk.
                  }
         \label{fig:fluxfunction_b}         
         \end{subfigure}
         \caption{Shown is the flux function $f$ from equations~(\ref{eq:f_pagethorne}) 
                  and (\ref{eq:f_pagethorne_mod}) for different values of $a$ (and $B$). If not
                  stated otherwise $B=(64/27)m^3$ is assumed for the pc-GR case.
                  }
    \end{figure}%
    For small values of $a$ we still have an ISCO in the pc-GR case and the flux looks similar 
    to the standard GR flux -- it is comparable to standard GR with higher values of $a$. If $a$
    increases and we do not have a last stable orbit in the pc-GR case, the flux gets 
    significantly larger and now has a minimum. This minimum can be seen as a dark ring 
    in the accretion disks in Fig.~\ref{fig:accretiondisk_1}.

    Another feature is the change of shape of the higher order images. For spin values 
    of $a\geq 0.416m$ the disk extends up to the central object in the pc-GR model, as it is
    the case for (nearly) extreme spinning objects in standard GR. Therefore no higher order images
    can be seen in this case. However in Figs.~\ref{fig:ptdisk_std_0}-\ref{fig:ptdisk_std_09},
    \ref{fig:disk_pc_a0} and \ref{fig:disk_pc_a03} images of higher order occur. The ringlike
    shapes in Figs.~\ref{fig:disk_pc_a06} and~\ref{fig:disk_pc_a06} are not images of higher
    order but still parts of the original disk, as described above. They could be mistaken for
    images of higher order although they differ significantly on the redshifted side of the disk.

\clearpage
    \begin{figure}
        \centering
        \begin{subfigure}{0.47\columnwidth}
                \centering
                \includegraphics[width=\textwidth]{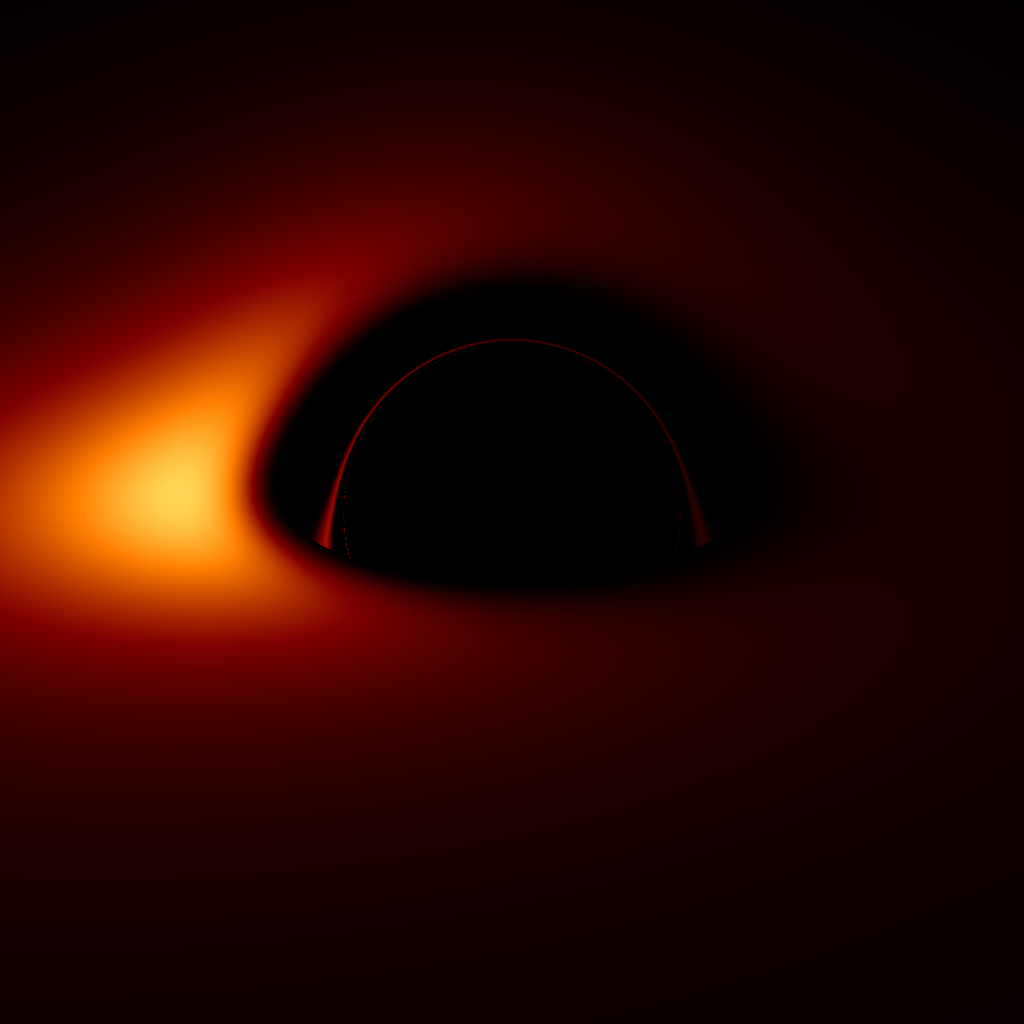}
                \caption{standard GR $a=0.0m$}
                \label{fig:ptdisk_std_0}
        \end{subfigure}%
        \qquad
        \begin{subfigure}{0.47\columnwidth}
                \centering
                \includegraphics[width=\textwidth]{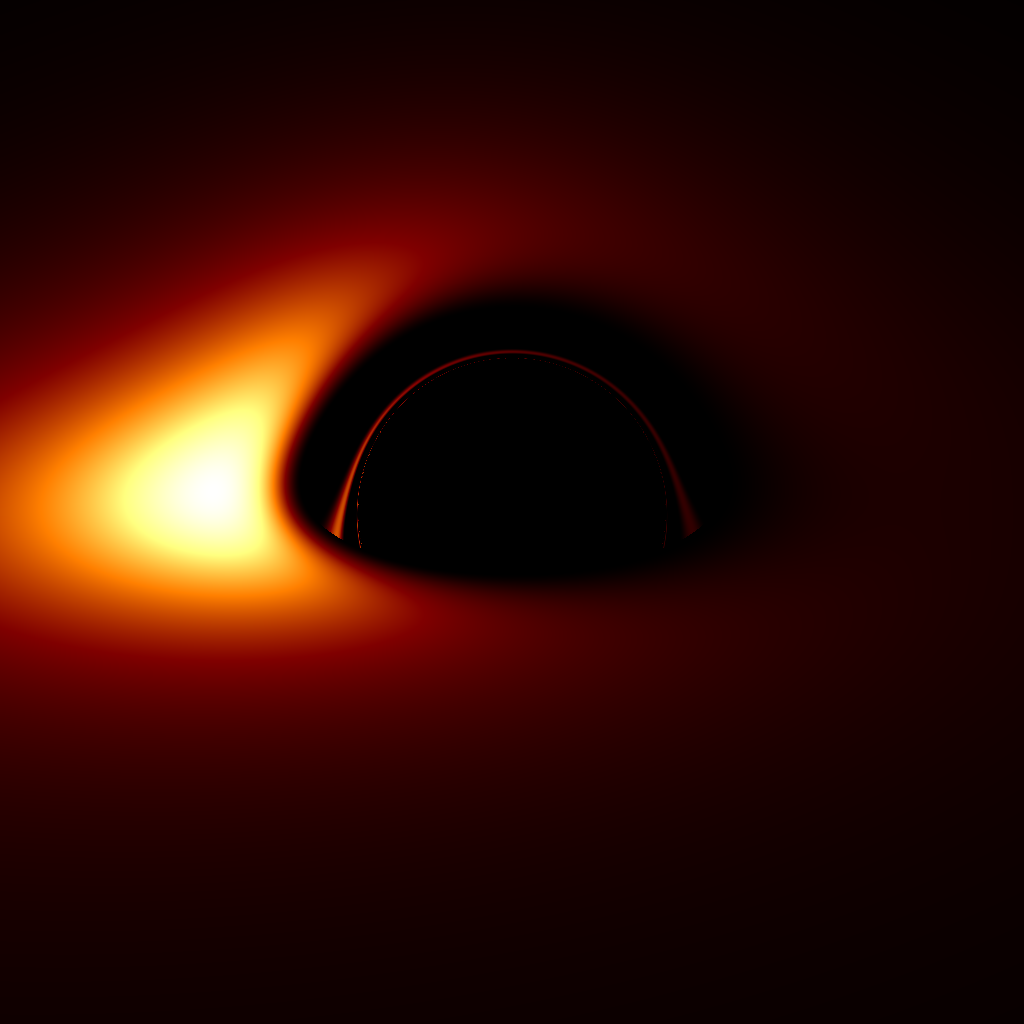}
                \caption{pc-GR $a=0.0m$}
                \label{fig:disk_pc_a0}
        \end{subfigure}%
                
        \begin{subfigure}{0.47\columnwidth}
                \centering
                \includegraphics[width=\textwidth]{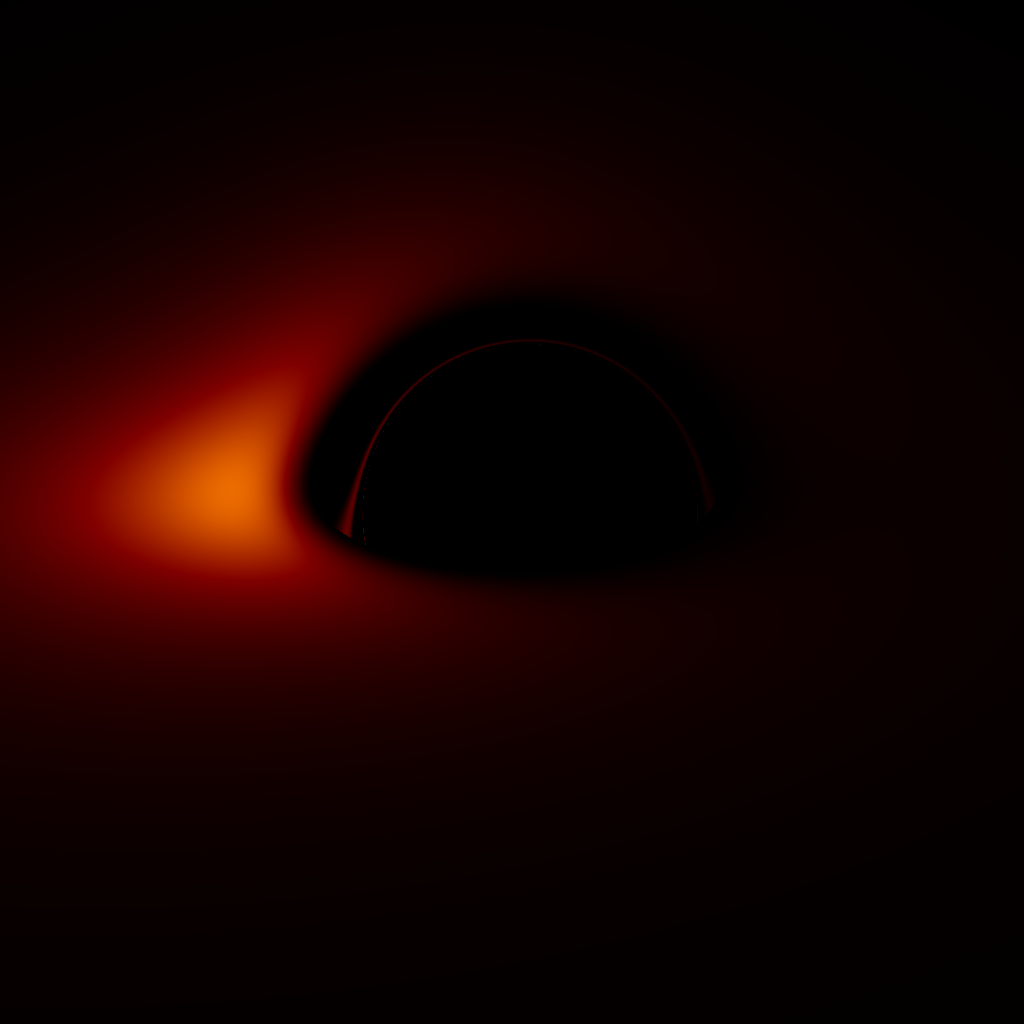}
                \caption{standard GR $a=0.3m$}
        \end{subfigure}%
        \qquad
        \begin{subfigure}{0.47\columnwidth}
                \centering
                \includegraphics[width=\textwidth]{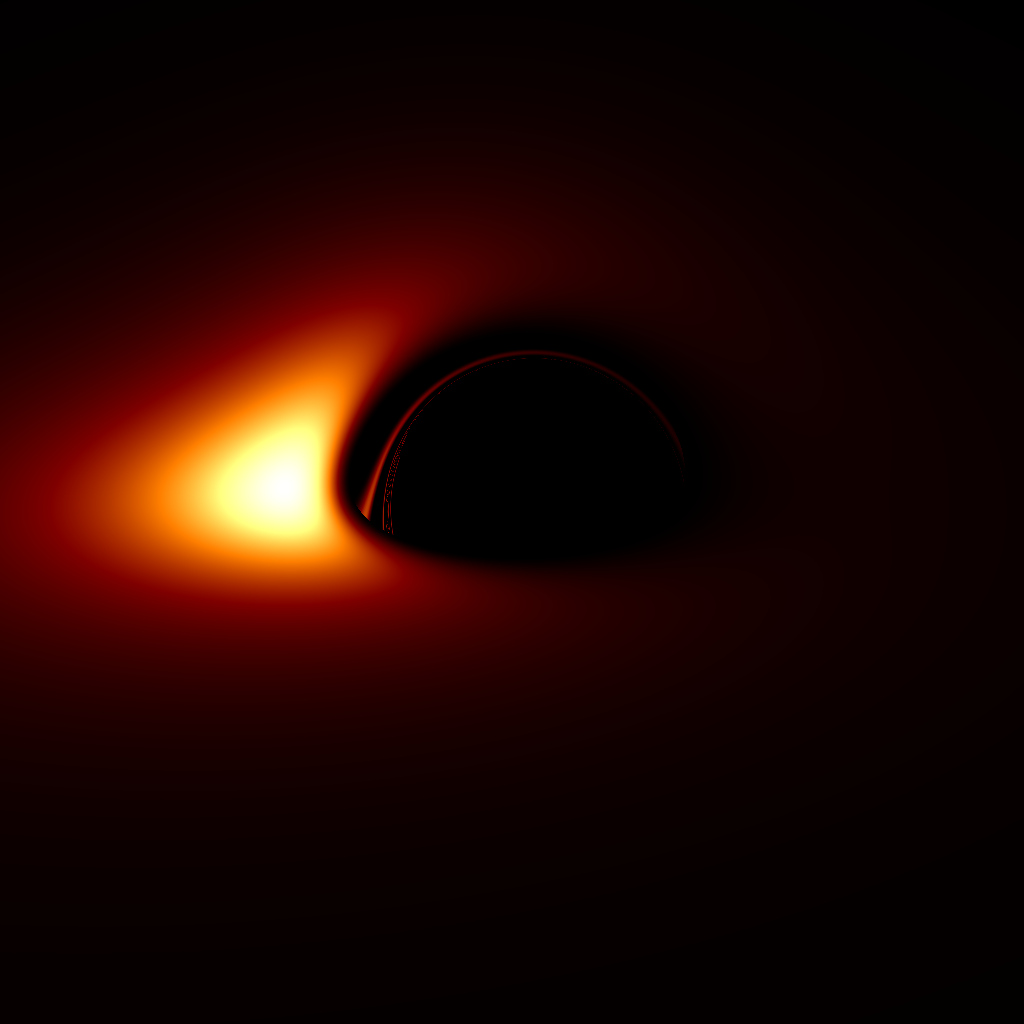}
                \caption{pc-GR $a=0.3m$}
                \label{fig:disk_pc_a03}        
        \end{subfigure}%
        \phantomcaption
    \end{figure}
    
    \begin{figure}
    \ContinuedFloat
        \centering
        \begin{subfigure}{0.47\columnwidth}
                \centering
                \includegraphics[width=\textwidth]{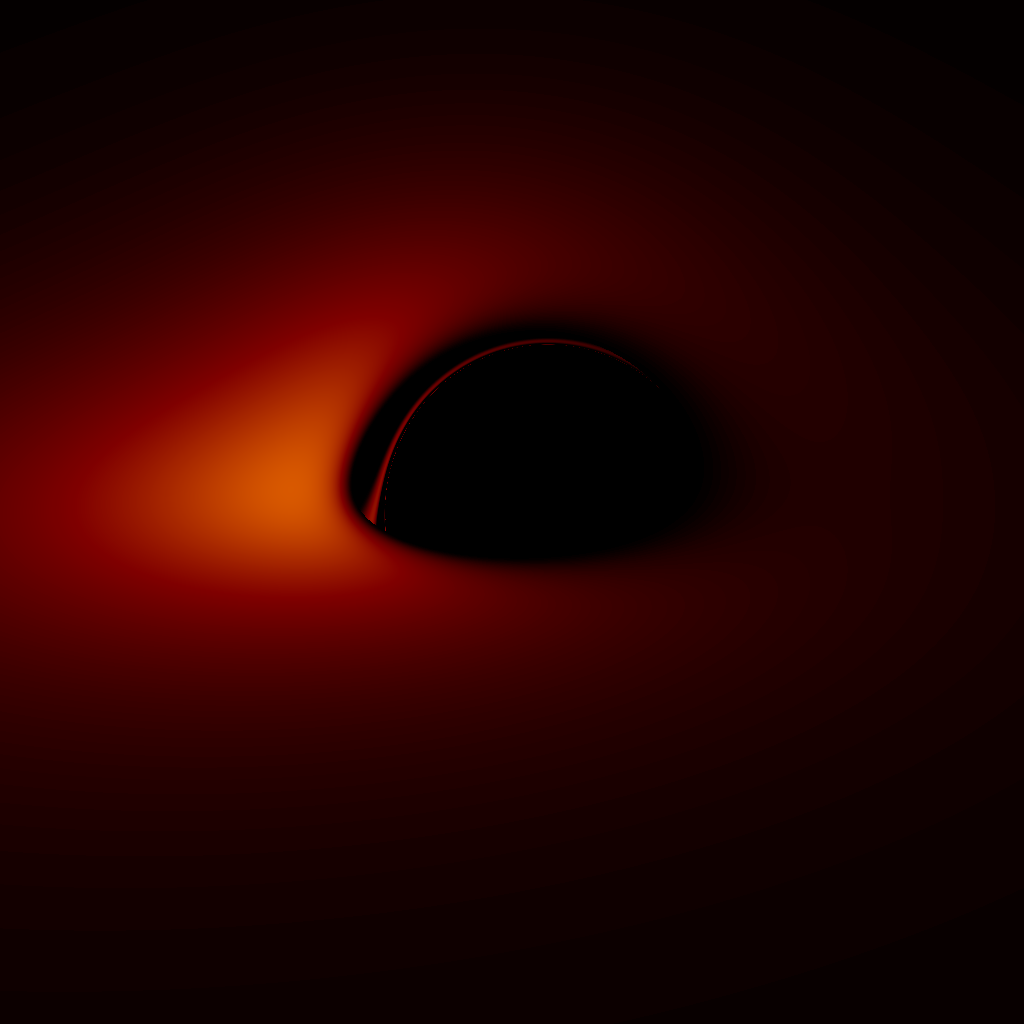}
                \caption{standard GR $a=0.6m$}     
        \end{subfigure}%
        \qquad
        \begin{subfigure}{0.47\columnwidth}
                \centering
                \includegraphics[width=\textwidth]{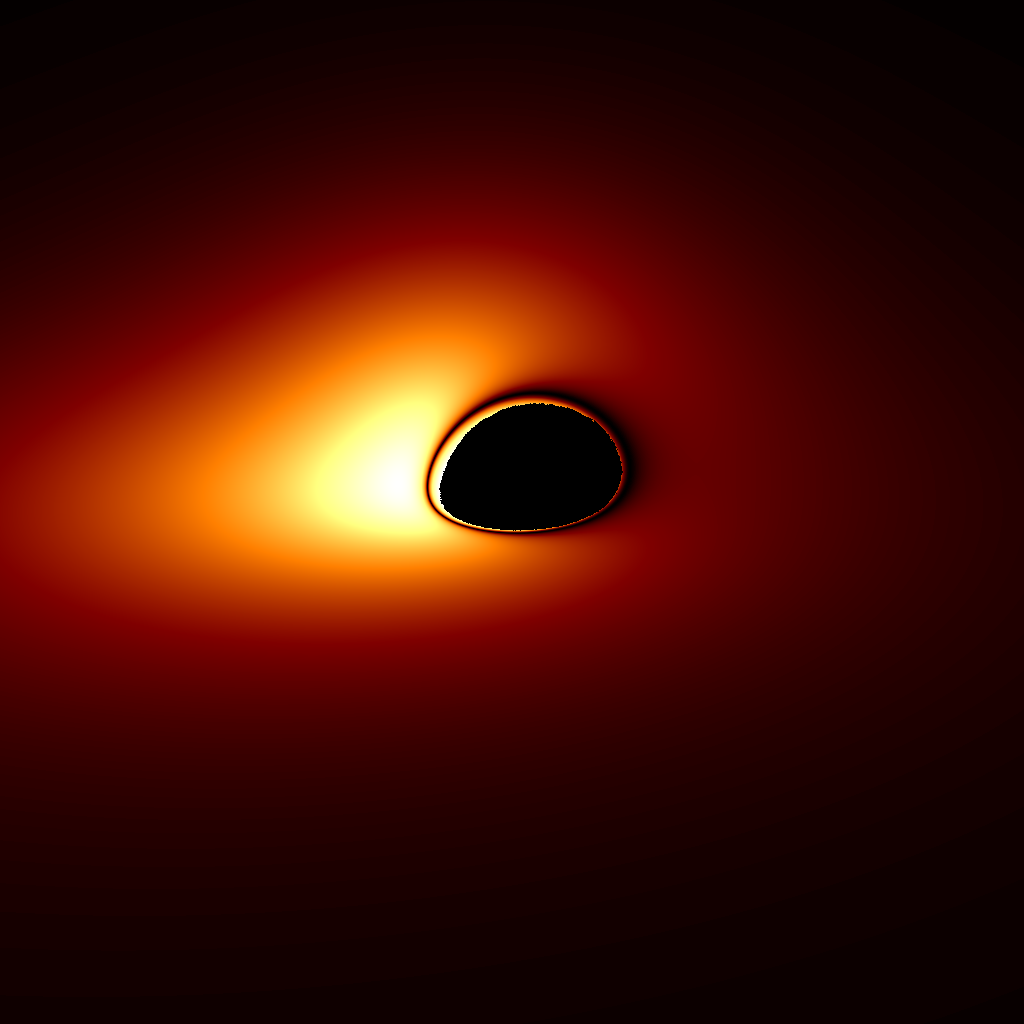}
                \caption{pc-GR $a=0.6m$}
                \label{fig:disk_pc_a06}
        \end{subfigure}%

        \begin{subfigure}{0.47\columnwidth}
                \centering
                \includegraphics[width=\textwidth]{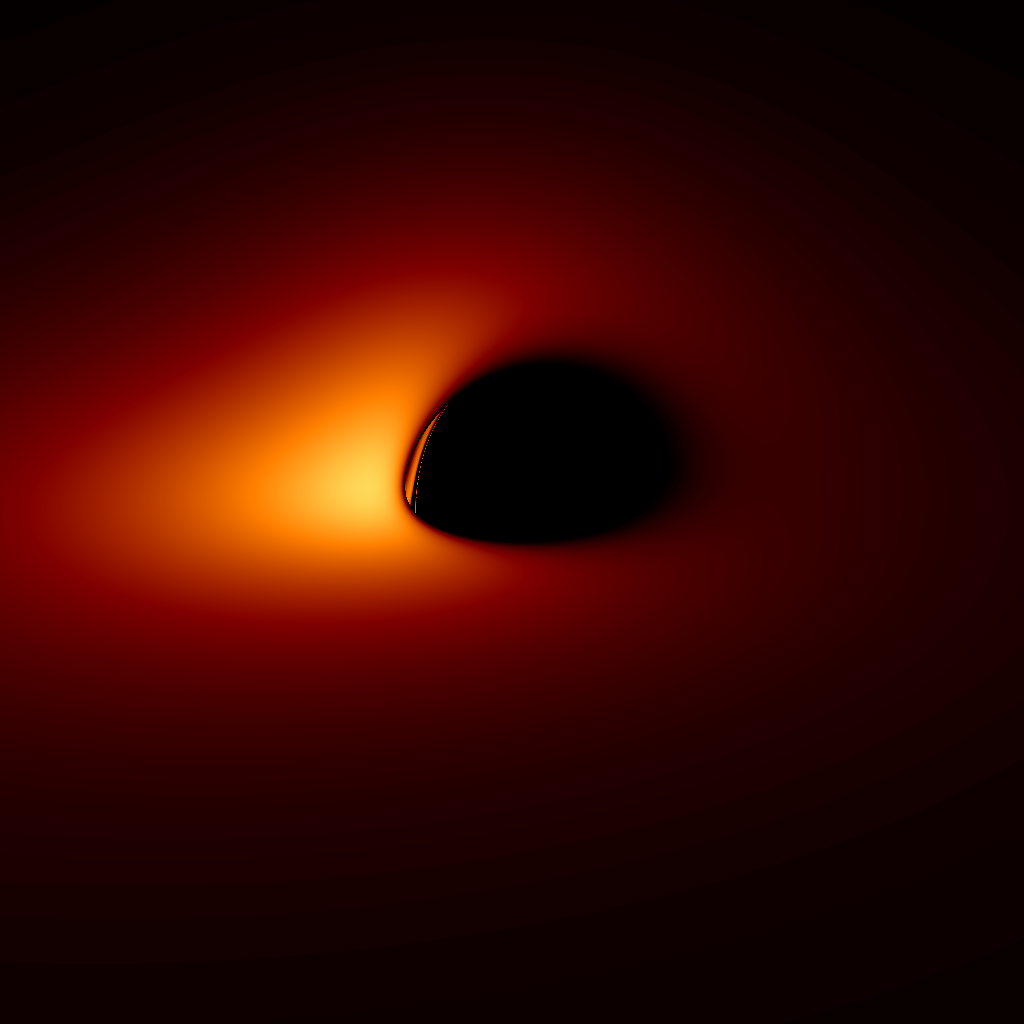}
                \caption{standard GR $a=0.9m$}     
                \label{fig:ptdisk_std_09}
        \end{subfigure}%
        \qquad
        \begin{subfigure}{0.47\columnwidth}
                \centering
                \includegraphics[width=\textwidth]{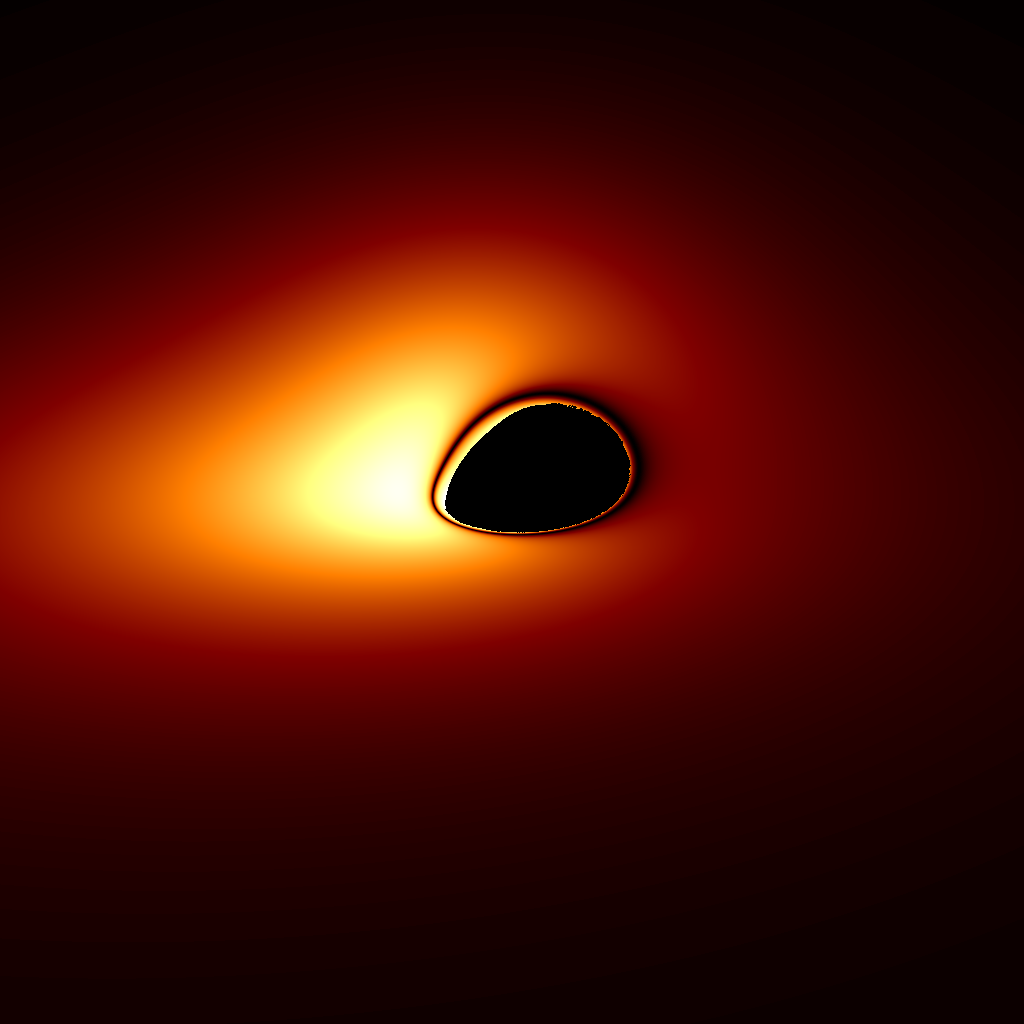}
                \caption{pc-GR $a=0.9m$}
                \label{fig:disk_pc_a09}
        \end{subfigure}%
        \caption{Infinite, counter clockwisely geometrically thin
                 accretion disk around static and rotating compact
                 objects viewed from an inclination of 70\textdegree.
                 The left panel shows the original disk model by
                 \protect\cite{pagethorne}. The right panel shows the
                 modified model, including pc-GR correction terms as
                 described in section \protect\ref{sec:observables}.
                 Scales change between the images.}
        \label{fig:accretiondisk_1}
    \end{figure}
\clearpage
  \subsection{Emission line profiles for the iron K\texorpdfstring{$\alpha$}{TEXT} line}
    As mentioned earlier, emission line profiles allow to investigate regions of strong gravity. 
    All results in this section share the same parameter values for the outer radius of the disk
    ($r = 100m$), the inclination angle ($\theta = 40\degree$) and the power law parameter 
    $\alpha=3$ (as suggested for disks first modelled by \cite{Shakura:1972}), see 
    equation~(\ref{eq:powerlaw_line}). We use this simpler model to simulate emission lines
    as it is widely used in the literature and thus results are easily comparable.     
    The angle of $\theta = 40\degree$ is just 
    an exemplary value and can be adjusted. As rest energy for the iron K$\alpha$ line
    we use $6.4$ keV. The inner radius of the disks is determined by the ISCO in
    GR and by the values in Tab.~\ref{tab:rin_pc} for pc-GR, and varies with varying values for 
    $a$. Shown is the flux in arbitrary units.
    In Fig.~\ref{fig:line_some_pcGR} and \ref{fig:line_some_stdGR} we compare 
    the influence of the objects spin on the shape of the emission line profile in GR and pc-GR 
    separately. Both in GR and pc-GR we observe the characteristic broad and smeared out low energy 
    tail, which grows with growing spin. It is more prominent in the case of pc-GR.
    The overall behaviour is the same in both theories. 
    \begin{figure}
      \centering
      \begin{subfigure}{\columnwidth}
         \centering
         \includegraphics[width=.79\textwidth]{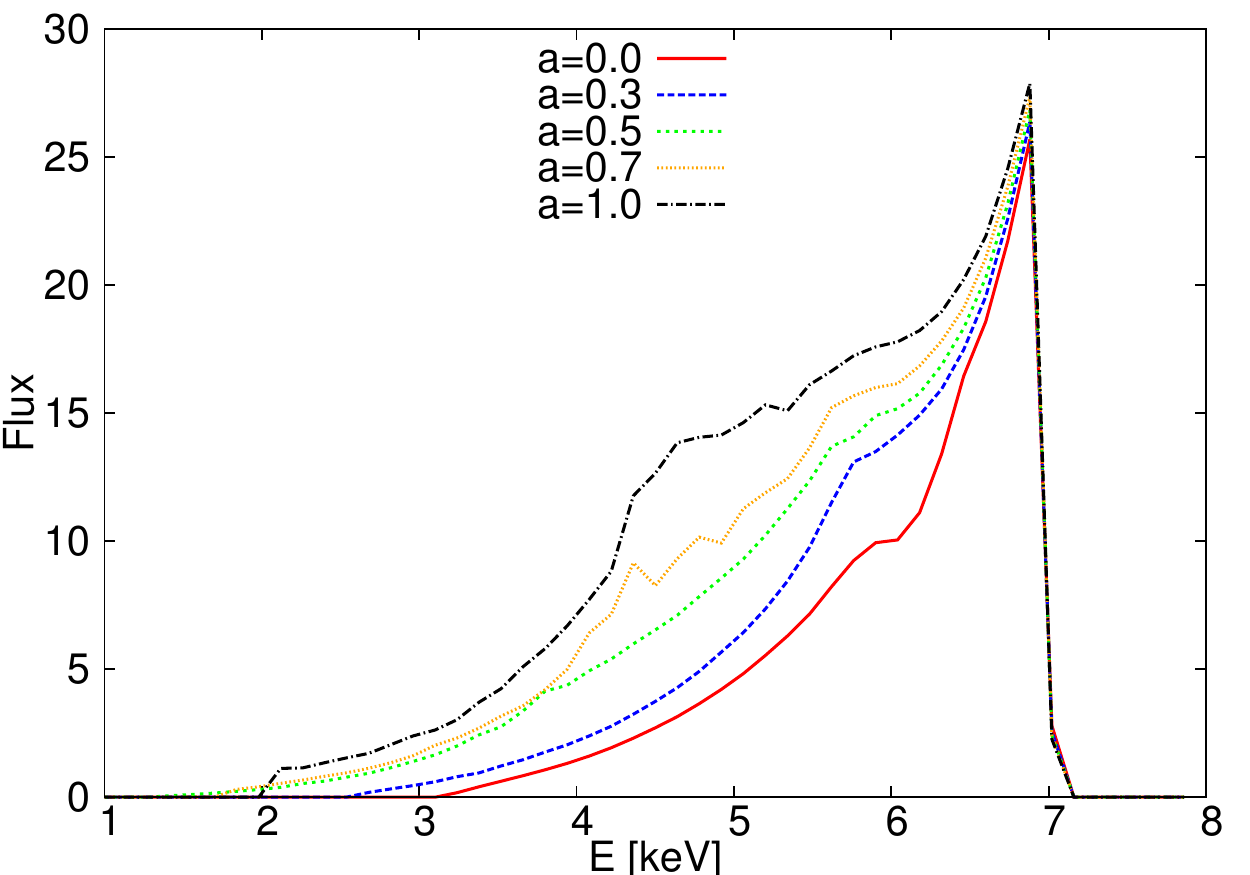}
         \caption{pc-GR}
         \label{fig:line_some_pcGR}
      \end{subfigure}
      \begin{subfigure}{\columnwidth}
        \centering
        \includegraphics[width=.79\textwidth]{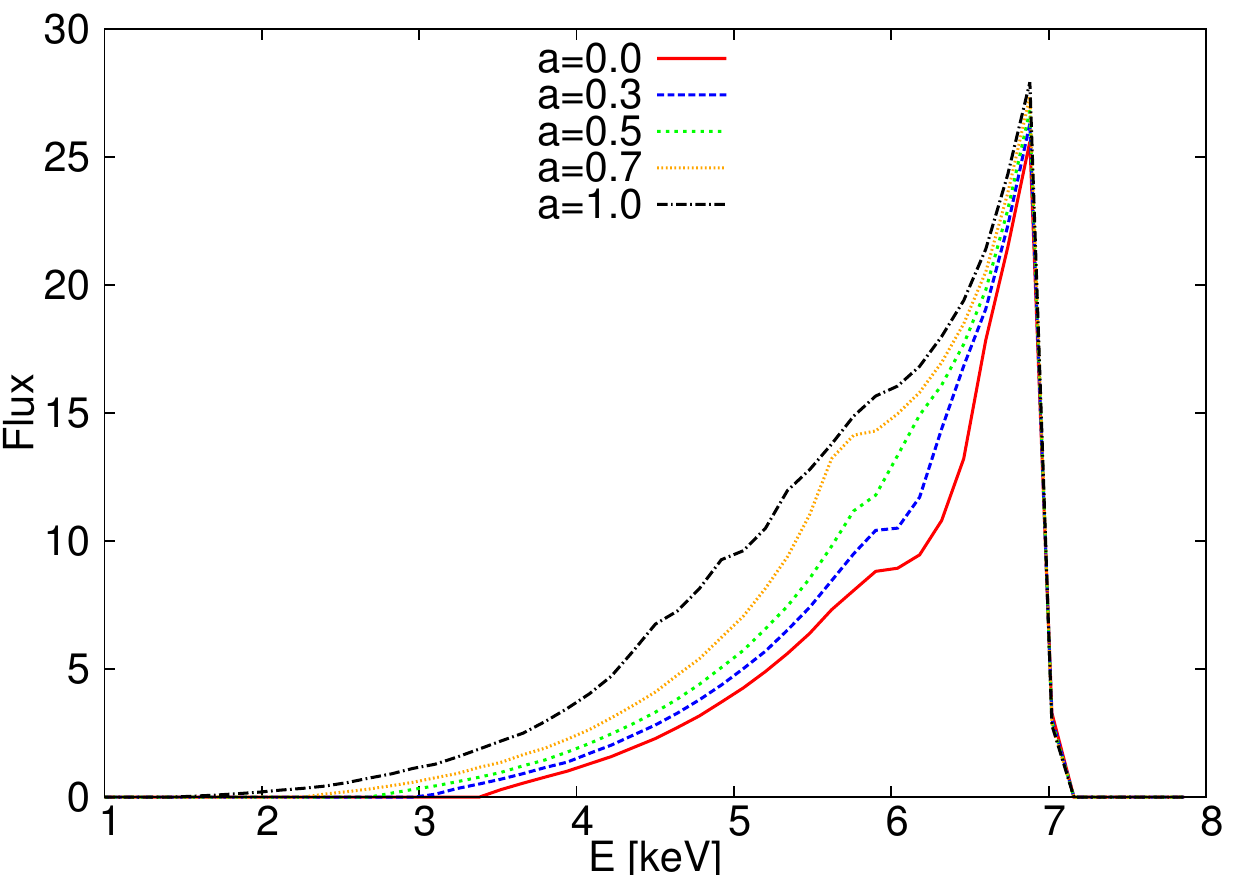}
        \caption{standard GR}
        \label{fig:line_some_stdGR}       
      \end{subfigure}
      \caption{Several line profiles for different values of 
               the spin parameter $a$.}
    \end{figure}%
    A closer comparison of both theories and their differences is then done in 
    Figs.~\ref{fig:lines_both_1} and \ref{fig:lines_both_2}, where we compare the two theories for different values of the
    spin parameter $a$. For slow rotating objects (Schwarzschild limit), almost no difference is
    observable. As the spin grows, we observe an increase of the low energy tail in the 
    pc-GR scenario compared to the GR one. The blue shifted peak however stays nearly the same.
    \begin{figure}
        \centering
        \begin{subfigure}{\columnwidth}
                \centering
                \includegraphics[width=.79\textwidth]{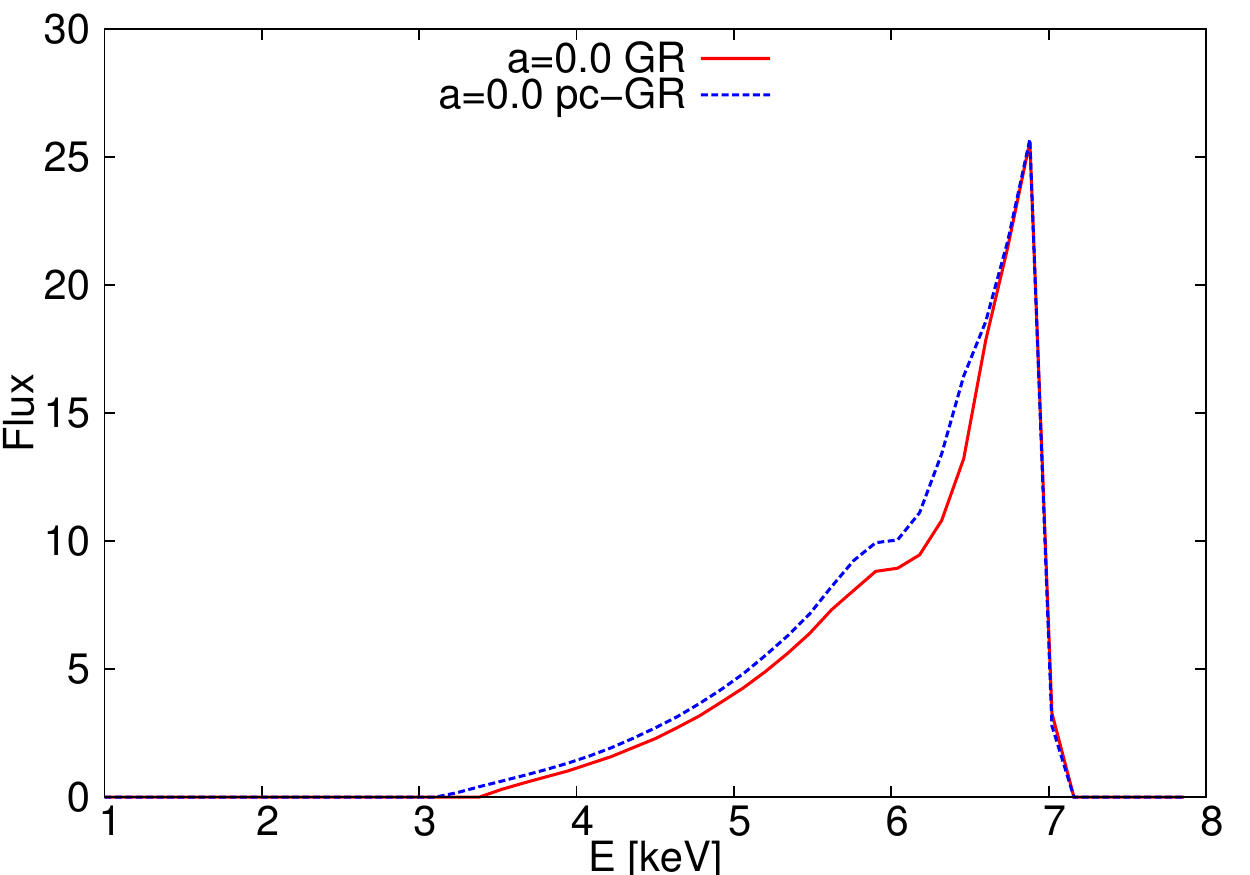}
                \caption{$a=0.0m$}
                \label{fig:line_both_00}
        \end{subfigure}%

        \begin{subfigure}{\columnwidth}
                \centering
                \includegraphics[width=.79\textwidth]{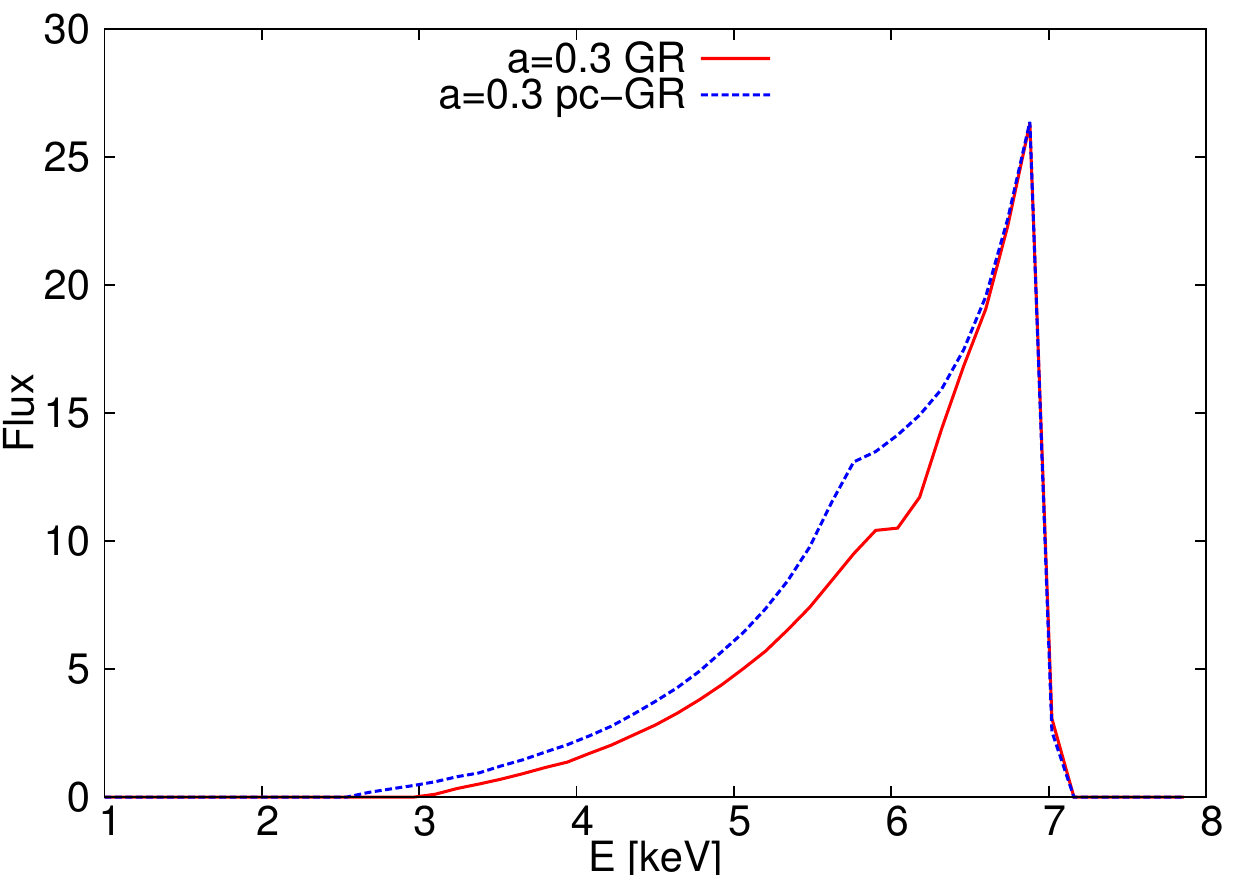}
                \caption{$a=0.3m$}
        \end{subfigure}%
        \caption{Comparison between theories. The plots are done for parameter values $r = 100m$ 
                 for  the outer radius of the disk, $\theta = 40\degree$ for the inclination angle
                 and $\alpha=3$ for the power law parameter. The inner radius of the disks is 
                 determined by the ISCO and thus varies for varying $a$.
                 }
        \label{fig:lines_both_1}
    \end{figure}
    
    \begin{figure}
      \centering
      \begin{subfigure}{\columnwidth}
        \centering
        \includegraphics[width=.79\textwidth]{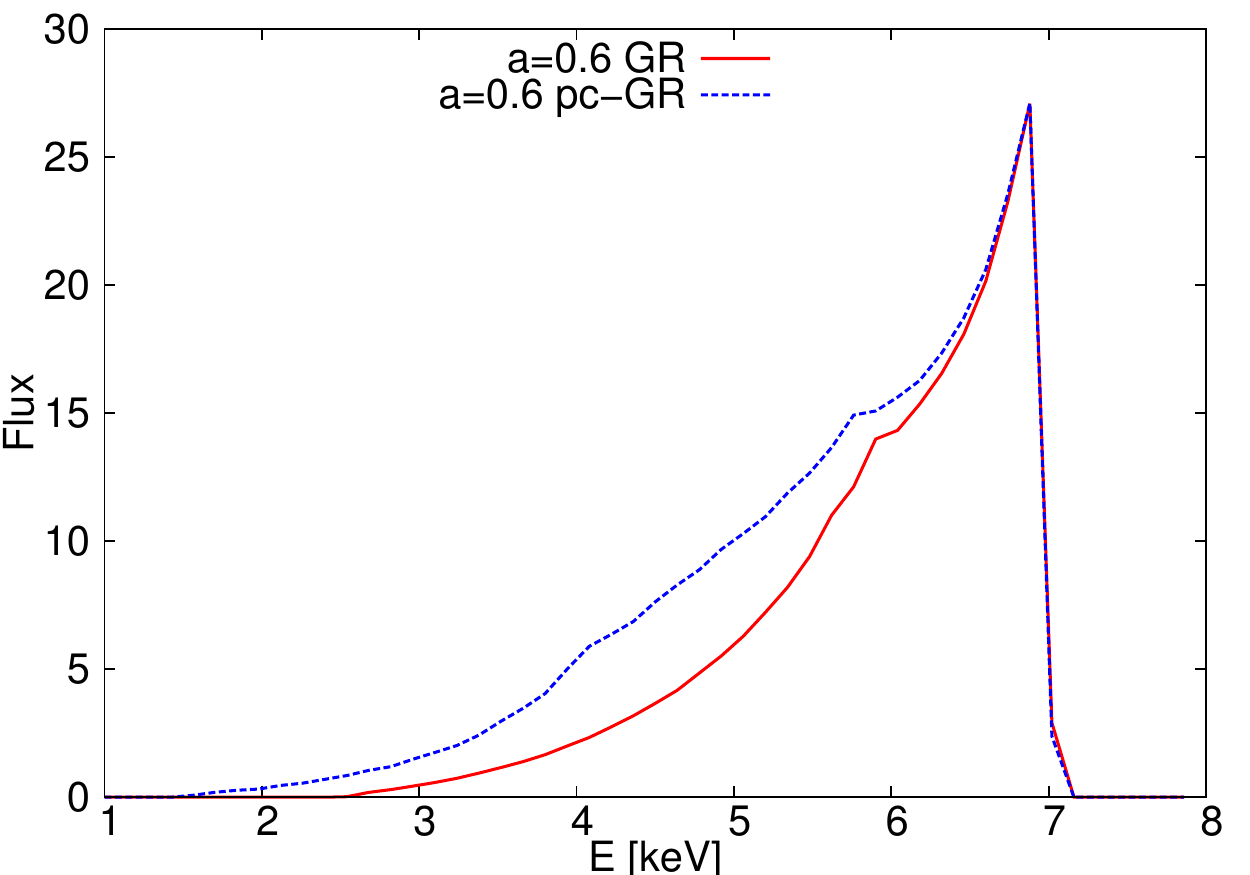}
        \caption{$a=0.6m$}
      \end{subfigure}%

      \begin{subfigure}{\columnwidth}
        \centering
        \includegraphics[width=.79\textwidth]{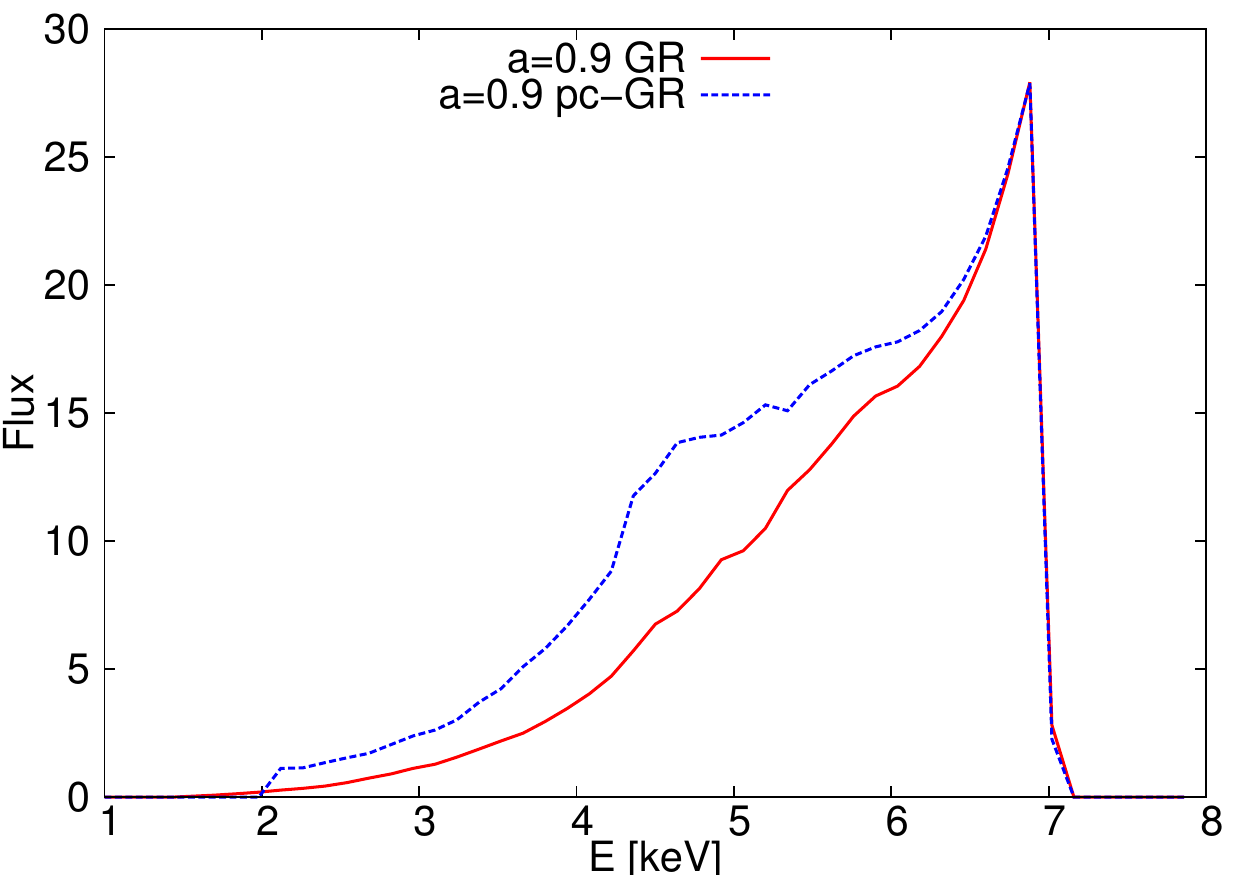}
        \caption{$a=0.9m$}
      \end{subfigure}%
      \caption{Comparison between theories. The plots are done for parameter values $r = 100m$ 
                 for  the outer radius of the disk, $\theta = 40\degree$ for the inclination angle
                 and $\alpha=3$ for the power law parameter. The inner radius of the disks is 
                 determined by the ISCO and thus varies for varying $a$.
                 }
      \label{fig:lines_both_2}
    \end{figure}
    If we compare both theories for different values of the spin parameter $a$ they get almost
    indistinguishable for certain choices of parameters, see Fig.~\ref{fig:different_a}.
    \begin{figure}
      \centering
      \includegraphics[width=\columnwidth]{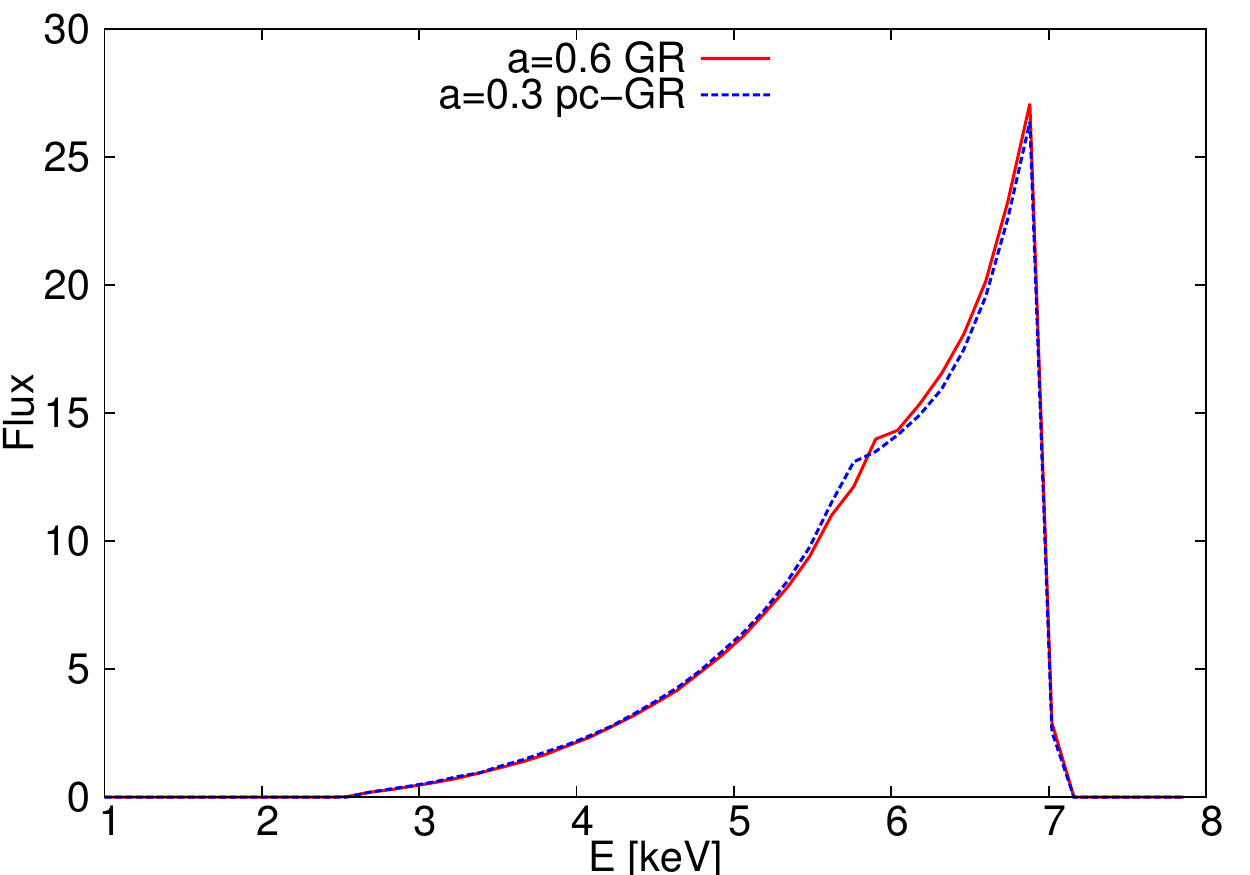}
      \caption{Comparison between theories for different values for the spin parameter. The plot is
               done for parameter values $r = 100m$ for  the outer radius of the disk, 
               $\theta = 40\degree$ for the inclination angle and $\alpha=3$ for the power 
               law parameter. The inner radius of the disks is determined by the ISCO.}
      \label{fig:different_a}
    \end{figure}

    To better understand the emission line profiles we have a look at the redshift in 
    two ways. The redshift can be written as \citep{fanton}
    \begin{equation}
     g= \frac{1}{u^0_\text{em}(1 - \omega \lambda)}  \quad,
    \end{equation}
    where $u^0_\text{em} = \frac{1}{\sqrt{-g_{00} - 2 \omega g_{03} - \omega^2 g_{33}}}$ is the 
    time component of the emitters four-velocity, $\omega$
    is the angular frequency of the emitter and $\lambda$ is the ratio of the emitted photons 
    energy to angular momentum.
    \cite{cisneros} derived an expression for photons emitted directly in the direction of the emitters
    movement
    \begin{equation}
      \lambda_\text{cis} = \frac{-g_{03} - \sqrt{g_{03}^2 - g_{00} g_{33}}}{g_{00}} 
    \end{equation}
    We take this expression and use it to approximate the redshift viewed from an inclination
    angle $\theta_\text{obs}$ as
    \begin{equation}
      g \approx \frac{1}{u^0_\text{em}(1 - \omega \lambda_\text{cis} \sin\theta_\text{obs})}  
    \end{equation}
    In Fig.~\ref{fig:combined_doppler_redshift} we show plots for different values of the spin 
    parameter for both GR and the pc-GR model for particles moving towards the observer, where we
    expect the highest blueshift to occur. To obtain the full frequency shift one needs in general
    to know the emission angle of the photon at the point of emission, which can be obtained by 
    using ray-tracing techniques. In Fig.~\ref{fig:redshift_thindisks} we display this 
    redshift obtained with GYOTO for a thin disk. Several features can now be seen in Figs.~\ref{fig:combined_doppler_redshift} and \ref{fig:redshift_thindisks}. First we see that 
    the maximal blueshift is almost the same in both the GR and pc-GR case. Then as the disks 
    extend to smaller radii in pc-GR we observe that there is a region where photons get redshifted,
    which is not accessible in GR for the same values of the spin parameter $a$. This can explain 
    the excess of flux in the redshifted region seen in Figs.~\ref{fig:lines_both_1} and 
    \ref{fig:lines_both_2} even for low values of the spin parameter. Finally the similarity of both theories
    for different values of parameters as shown in Fig.~\ref{fig:different_a} can also be seen
    in Figs.~\ref{fig:red_std_06} and \ref{fig:red_pc_03}.
        \begin{figure}
          \centering
          \begin{subfigure}{\columnwidth}
            \centering
            \includegraphics[width=.79\textwidth]{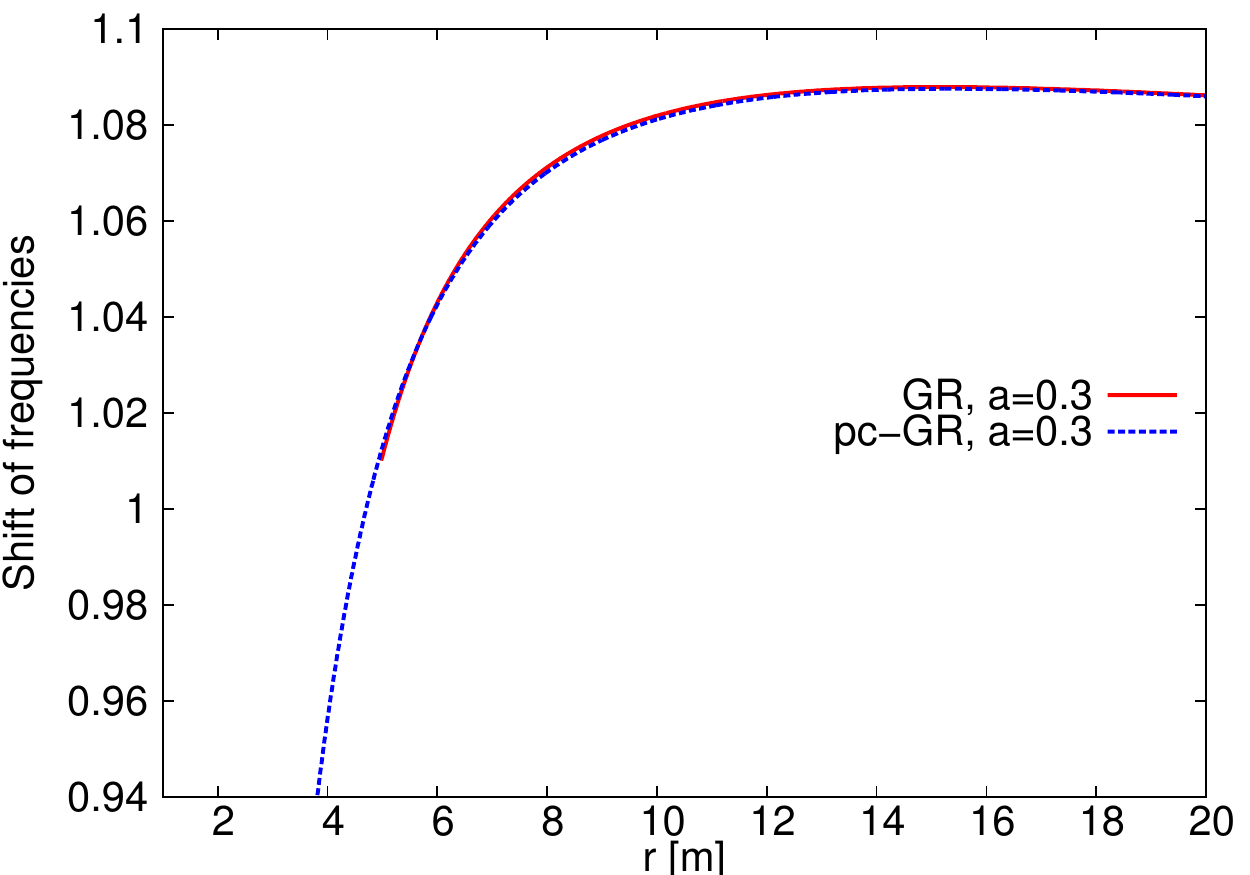}
            \caption{$a=0.3m$}
          \end{subfigure}%

          \begin{subfigure}{\columnwidth}
            \centering
            \includegraphics[width=.79\textwidth]{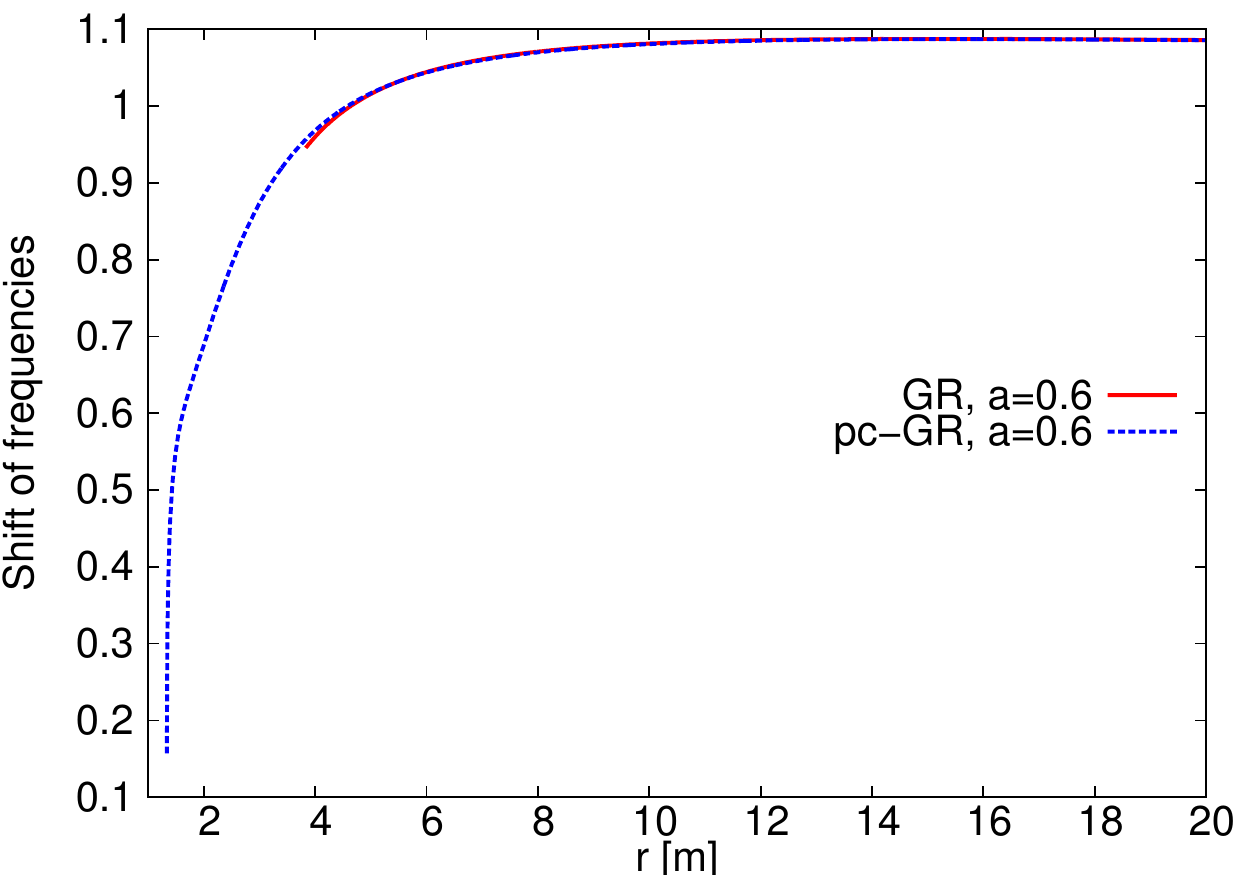}
            \caption{$a=0.6m$}
          \end{subfigure}%
          \caption{Combined effects of relativistic Doppler blueshift and gravitational
                   redshift as a function of the radius. The inclination is given as 
                   $\theta=40\degree$. Values greater than 1 represent a blueshift. The plots
                   start at the inner edge of the disk and are done for photons
                   emitted parallel to the direction of movement
                   of the emitter, i.e. where the highest blueshift occurs.}
          \label{fig:combined_doppler_redshift}
        \end{figure}%

    \begin{figure}
        \centering
        \begin{subfigure}{0.47\columnwidth}
                \centering
                \includegraphics[width=\textwidth]{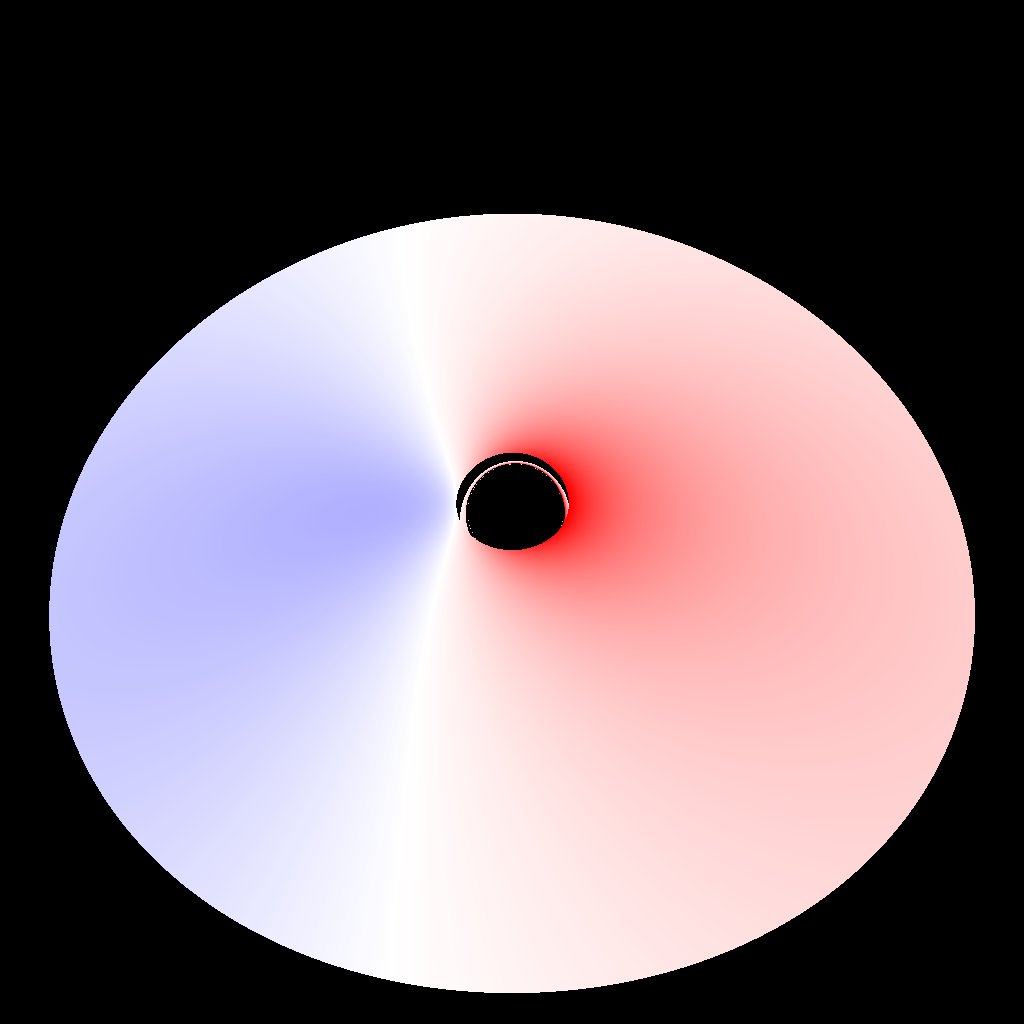}
                \caption{standard GR $a=0.3m$}
        \end{subfigure}%
        \qquad
        \begin{subfigure}{0.47\columnwidth}
                \centering
                \includegraphics[width=\textwidth]{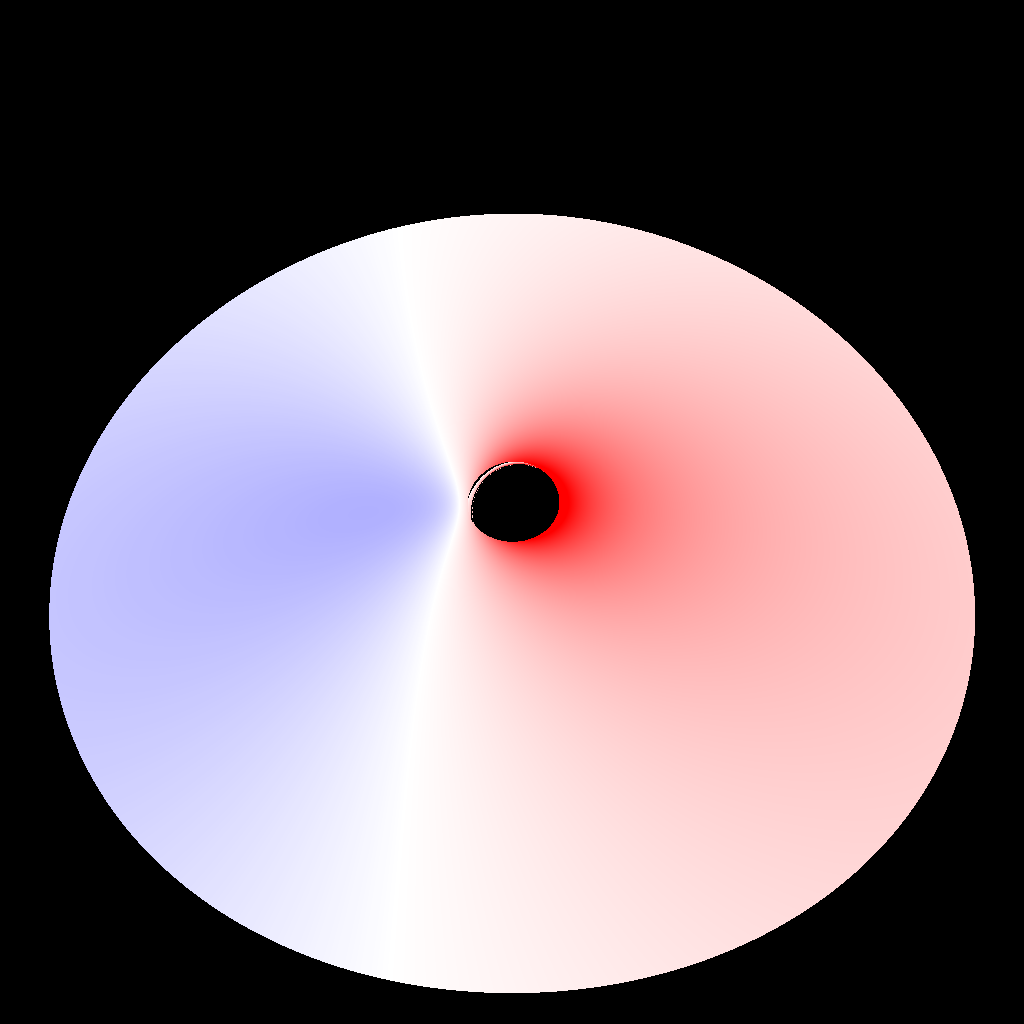}
                \caption{standard GR $a=0.6m$}
                \label{fig:red_std_06}
        \end{subfigure}%

        \begin{subfigure}{0.47\columnwidth}
                \centering
                \includegraphics[width=\textwidth]{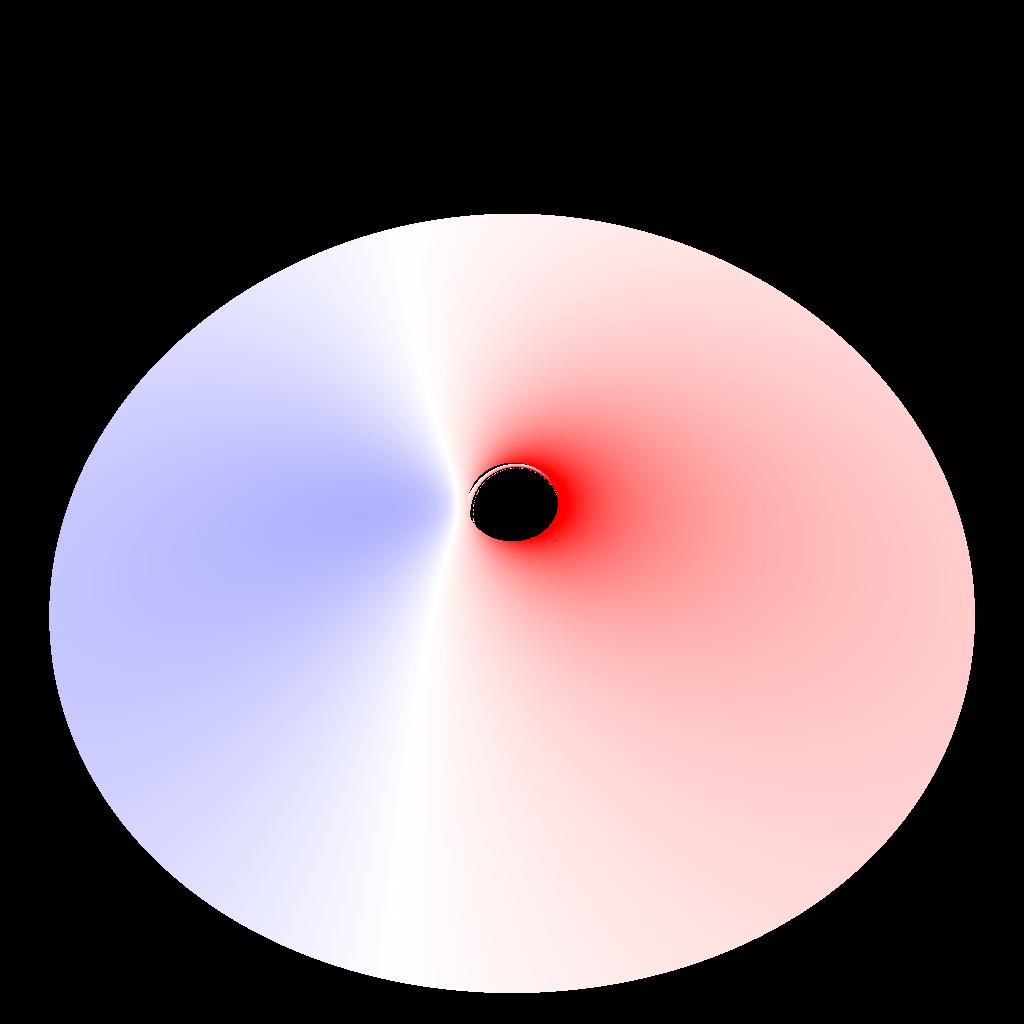}
                \caption{pc-GR $a=0.3m$}     
                \label{fig:red_pc_03}
        \end{subfigure}%
        \qquad
        \begin{subfigure}{0.47\columnwidth}
                \centering
                \includegraphics[width=\textwidth]{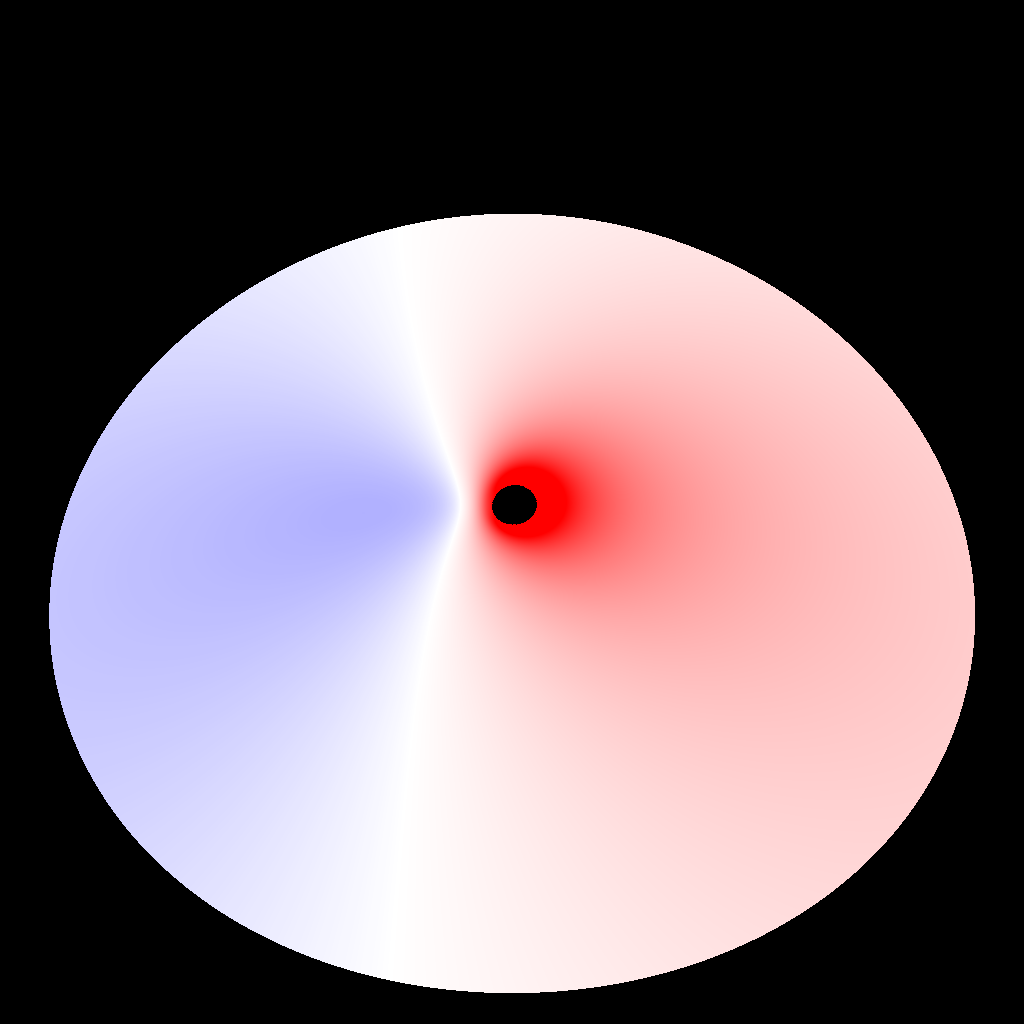}
                \caption{pc-GR $a=0.6m$}     
        \end{subfigure}%
        \caption{Redshift for a thin accretion disk. The inclination angle is 40\textdegree.
                 The outer radius is set to $r_\text{out}=50m$. The inner radius
                 is set to the ISCO, if exists. For the pc-GR case see Tab.~\ref{tab:rin_pc}. 
                 The similarity between the pc-GR case for $a=0.3m$ and standard GR for $a=0.6m$ 
                 can also be seen in Fig.~\ref{fig:different_a}.}
        \label{fig:redshift_thindisks}
    \end{figure}


\section{Conclusion}
  \label{sec:conclusion}
  We have adapted two models, which are implemented in \textsc{Gyoto} \citep{gyoto} -- an infinite, 
  geometrically thin and optically thick accretion disk \citep{pagethorne} and the iron K$\alpha$ 
  emission line profile of a geometrically thin and optically thick disk \citep{fanton} -- 
  to incorporate correction terms due to a pseudo-complex extension of GR.
  In both models we can see differences between standard GR and pc-GR. Those differences can be
  attributed to the modification of the last stable orbit in pc-GR and thus disks which extend 
  further in for a big range of spin parameter values of the massive object. In addition
  the gravitational redshift and orbital frequencies of test particles have to be modified. Both the
  accretion disk images and emission lines profiles show an increase in the amount of outgoing
  radiation thus turning the massive objects brighter in pc-GR than in GR, assuming that all other
  parameters are the same.
  Although the difference in the emission line profiles is in principle big enough to be used
  to discriminate between GR and pc-GR, the effects of the pc-correction terms on the results 
  are not as strong as the modifications presented, e.g. in \cite{bambi}. Also an uncertainty in, e.g.
  the spin parameter $a$ can make it very difficult to discriminate between both theories as we
  have seen in Fig.~\ref{fig:different_a}.  


\section*{Acknowledgements}
\label{sec:acknowledgments}
  The authors want to thank the referee for very detailed and valuable comments on this article.
  The authors express sincere gratitude for the possibility to work at
  the Frankfurt Institute of Advanced Studies with the excellent working
  atmosphere encountered there. The authors also want to thank the creators of \textsc{Gyoto} for
  making the programme open source, especially 
  Fr\'{e}d\'{e}ric Vincent for supplying them with an at that time unpublished version for 
  simulating emission line profiles.
  T.S expresses his gratitude for the possibility of a work stay at the Instituto de Ciencias
  Nucleares, UNAM. M.S. and T.S. acknowledge support from Stiftung Polytechnische 
  Gesellschaft Frankfurt am Main. P.O.H. acknowledges financial support from DGAPA-PAPIIT 
  (IN103212) and CONACyT. 
  G.C. acknowledges financial support from Frankfurt Institute for Advanced Studies.



\end{document}